\documentclass[letterpaper,11pt]{article}
\usepackage{jheppub}
\usepackage{comment}
\pdfoutput=1
\usepackage{dcolumn}% Align table columns on decimal point
\usepackage{bm}% bold math
\usepackage{graphicx}% Include figure files
\usepackage{diagbox}
\usepackage{amssymb,amsmath}
\usepackage{multirow}
\usepackage{color,url}
\usepackage{tabu}
\usepackage{array}
\usepackage{ulem}
\usepackage{soul}
\usepackage{colordvi}
\usepackage{subcaption}
\usepackage{ mathrsfs }
\def\ga{\mathrel{\raise.3ex\hbox{$$>$$\kern-.75em\lower1ex\hbox{$\sim$}}}}
\def\la{\mathrel{\raise.3ex\hbox{$<$\kern-.75em\lower1ex\hbox{$\sim$}}}}
\def\beqa{\begin{eqnarray}}
\def\eeqa{\end{eqnarray}}

\begin{document}

\title{\boldmath Sensitivity on Two-Higgs-Doublet Models from Higgs-Pair Production via $b\bar{b}b\bar{b}$ Final State}

\author[1]{Yi-Lun Chung,}
\author[1,3]{Kingman Cheung,}
\author[2]{and Shih-Chieh Hsu}
\affiliation[1]{\normalsize Department of Physics and Center for Theory and Computation, National Tsing Hua University, Hsinchu 300, Taiwan}
\affiliation[2]{\normalsize Department of Physics, University of Washington, Seattle, Washington 98195, USA}
\affiliation[3]{\normalsize Division of Quantum Phases and Devices, 
School of Physics, Konkuk University, Seoul 143-701, Republic of Korea}

\emailAdd{cheung@phys.nthu.edu.tw}
\emailAdd{s107022801@m107.nthu.edu.tw}
\emailAdd{schsu@uw.edu}

\date{\today}

\abstract{
Higgs boson pair production is well known to probe the structure of the
electroweak symmetry breaking sector. We illustrate using the gluon-fusion process $pp \to H \to h h \to (b\bar b) (b\bar b)$ in the framework of 
two-Higgs-doublet models and how the machine learning approach (three-stream convolutional neural network) can substantially improve the signal-background
discrimination and thus improves the sensitivity coverage of the relevant
parameter space. We show that such $gg \to hh \to b \bar b b\bar b$ process can further probe the currently allowed parameter space by 
\textsc{HiggsSignals} and \textsc{HiggsBounds} at the HL-LHC. The results for Types I to IV are shown.
}

\maketitle

\newpage
% %%%%%%%%%%%%%%%%%%%%%%%%%%%%%%%%%%%%%%%%%%%%%%%%%%%%%%%%%%%%%%%%%%%%%%%%%%%%%%%%
% Introduction 
\section{Introduction}
Origin of mass is highly related to the mechanism involved in
electroweak symmetry breaking (EWSB), which is believed to give
masses to matters and gauge bosons.
The simplest implementation of EWSB in the standard model (SM) is to introduce 
a Higgs doublet field  \cite{Higgs:1964pj,Englert:1964et,Guralnik:1964eu}.
A neutral scalar Higgs boson was  then discovered in
July 2012 \cite{ATLAS:2012yve,CMS:2012qbp}, which is believed to serve as
the role of EWSB. 
After all the data accumulated till 2018, the scalar boson
is best described by the SM Higgs boson \cite{Cheung:2013kla,Cheung:2018ave}.
However, the SM Higgs boson cannot be a complete theory because of
the gauge hierarchy problem.  

There is no {\it a priori} reason why the EWSB sector simply contains only one 
Higgs doublet field. Indeed, many extensions of the EWSB sector consist of
more Higgs fields. One of the best ways to probe the structure of the Higgs sector is
to probe the Higgs self-couplings.  This is
because the self-couplings of the Higgs boson 
are very different among the SM, two-Higgs doublet models (2HDM), MSSM, and
any composite Higgs models.
One of the probes of Higgs self-couplings is Higgs-pair production via gluon fusion
at the LHC~\cite{Glover:1987nx,Dicus:1987ic,Plehn:1996wb,Djouadi:1999rca,Dawson:1998py,Baur:2002qd,Binoth:2006ym,Baur:2003gpa,Baglio:2012np,Grigo:2013rya,Barger:2013jfa}.
%,Baur:2003g
There have been a large number of works in literature on Higgs-pair production
beyond the SM (see for example \cite{Lu:2015jza} and references therein).

The predictions for various models are quite different such that the
production rates can give valuable information on the self-coupling $\lambda_{3H}$
or on the presence of heavier Higgs bosons which can enhance the production
rates for Higgs boson pairs.
The Higgs-pair production process receives contributions from both
the triangle and box diagrams, which interfere with each other.
In 2HDM, the triangle diagram can involve the Higgs self-trilinear coupling
$\lambda_{Hhh}$ and $\lambda_{hhh}$.  In particular, the resonance effect of the
heavier CP-even Higgs boson can substantially enhance the production rate of $hh$ pairs.
The production rate largely depends on the parameters of the 2HDM, such as
$\cos(\beta-\alpha)$, $\tan \beta$, and $m_{12}^2$ in addition to the $M_H$ and
$\Gamma_H$. 

In this work, we study the signal process
$pp \to  h h \to (b\bar b) (b\bar b)$ via gluon fusion against the SM multijet background.
It is well-known that the signal is overwhelmingly buried under the
multijet background. The study using the conventional cut-based approach
did not give enough significance even at the High-Luminosity LHC (HL-LHC).
We make use of the boosted feature of the final-state Higgs boson pair 
$hh$, due to the
decay of the heavier CP-even Higgs boson.  A specific classifier was developed in
Ref.~\cite{Chung:2020ysf}, which can be employed to significantly enhance the
signal-background ratio.  We show that the Three-stream Convolutional Neural Network (3CNN) can
substantially improve the significance of the signal compared to the
Boosted Decision Tree method (BDT) and the conventional cut-based approach.
At the end of the analysis, we show the 95\% sensitivity coverage of the
parameter space of the 2HDM Type I, II, III, and IV.  
The current study focuses on the boosted region of the Higgs boson pair, 
for which the classifier that we employ is very effective in reducing the
SM multijet background. 
In literature, there are existing analyses in probing the 
Higgs self-coupling in the channels such as  
$hh \to  b\bar b \gamma\gamma$ \cite{Chang:2018uwu,Chang:2019ncg}, 
$hh \to  b\bar b  W W^*$ \cite{Kim:2018cxf,Papaefstathiou:2012qe},
$ hh \to b\bar b b\bar b$~\cite{Amacker:2020bmn}, and resonance 
search \cite{Adhikary:2018ise}.

The organization is as follows. 
In the next section, we briefly describe the 2HDM's and relevant parameters for
Higgs-pair production. 
In Sec. \ref{sec:Sample}, we describe the signal and background processes, including the
sample generation and event selections.
In Sec. \ref{sec:classifiers}, we introduce the machine-learning approaches including BDT and 3CNN.
In Sec. \ref{sec:Sensitivity}, we scan the parameter space of the 2HDM's for the sensitivity coverage at the HL-LHC, as well as the current restriction on the parameter space due to \textsc{HiggsSignals} and \textsc{HiggsBounds}. We conclude in Sec. \ref{sec:conclusion}.

\section{Two Higgs Doublet Models}\label{sec:THDM}
Two-Higgs doublet model is an extension of the SM by adding another 
complex Higgs doublet field, and so the Higgs sector consists of 
$\Phi_1$ and $\Phi_2$~\cite{Branco:2011iw}:
\begin{equation}
\Phi_i = \left( \begin{array}{c} w_i^+ \\[3pt]
\dfrac{v_i +  h_i + i \eta_i }{ \sqrt{2}}
\end{array} \right), \quad i=1,2,
\end{equation}
where $v_{1}$ and $v_2$ are the vacuum expectation values (VEV) 
of $\Phi_1$ and $\Phi_2$, respectively.
The ratio of these two VEV's is defined by  $\tan \beta \equiv v_2/v_1$.
The dangerous flavor-changing-neutral-currents (FCNC) 
at tree level are avoided by imposing a discrete $Z_2$ symmetry, 
under which $\Phi_1 \to \Phi_1$ and $\Phi_2 \to -\Phi_2$~\cite{Glashow:1976nt,Paschos:1976ay}.
The scalar potential with softly broken $Z_2$ and 
\textit{CP} invariance is
\begin{eqnarray}
\label{eq:VH}
V = && m^2 _{11} \Phi^\dagger _1 \Phi_1 + m^2 _{22} \Phi^\dagger _2 \Phi_2
-m^2 _{12} ( \Phi^\dagger_1 \Phi_2 + {\rm H.c.}) \\ \nonumber
&& + \frac{1}{2}\lambda_1 (\Phi^\dagger _1 \Phi_1)^2
+ \frac{1}{2}\lambda_2 (\Phi^\dagger _2 \Phi_2 )^2
+ \lambda_3 (\Phi^\dagger _1 \Phi_1) (\Phi^\dagger _2 \Phi_2)
+ \lambda_4 (\Phi^\dagger_1 \Phi_2 ) (\Phi^\dagger _2 \Phi_1) \\ \nonumber
&& + \frac{1}{2} \lambda_5
\left[
(\Phi^\dagger _1 \Phi_2 )^2 +  {\rm H.c.}
\right],
\end{eqnarray}
where the $m^2 _{12}$ term softly breaks the $Z_2$ symmetry.

The model has five physical Higgs bosons: a pair of CP-even scalar 
bosons $h$ and $H$, a CP-odd pseudoscalar $A$, and a pair of 
charged Higgs bosons $H^\pm$.
The masses of the physical Higgs bosons are related to the 
$\lambda$'s in the scalar potential, the mixing angle $\alpha$ of the
CP-even scalar bosons and $\beta$, given by Ref.~\cite{Song:2019aav},
\begin{eqnarray}
\label{eq:quartic}
\lambda_1  &=& \frac{1}{v^2 c_\beta^2}
\left[
c_\alpha^2 M_H^2 + s_\alpha^2 m_h^2 - t_\beta m_{12}^2 
\right], \\ \nonumber
\lambda_2 &=& \frac{1}{v^2 s_\beta^2}
\left[
s_\alpha^2 M_H^2+c_\alpha^2 m_h^2 - \frac{1}{t_\beta}m_{12}^2
\right], \\ \nonumber
\lambda_3 &=& 
\frac{1}{v^2}
\left[
2 M_{H^\pm}^2 + \frac{s_{2\alpha}}{s_{2\beta} } (M_H^2-m_h^2) -\frac{m_{12}^2}{s_\beta c_\beta }
\right], \\ \nonumber
\lambda_4 &=& 
\frac{1}{v^2}
\left[
M_A^2- 2 M_{H^\pm}^2 + \frac{m_{12}^2}{s_\beta c_\beta}
\right], \\ \nonumber 
\lambda_5 &=&
\frac{1}{v^2}
\left[
\frac{m_{12}^2}{s_\beta c_\beta}- M_{A}^2
\right], 
\end{eqnarray}
where 
$s_\alpha \equiv \sin\alpha$, 
$c_\alpha \equiv \cos \alpha$, 
$t_\beta \equiv \tan \beta$, etc.
The six free parameters of the 2HDM's are 
\begin{equation}
    \{ m_h,\; M_H, \; M_A,\; M_{H^\pm},\; t_\beta,\; c_{\beta-\alpha} \}.
\end{equation}
Note that we focus on the scenario in which 
the lighter CP-even scalar Higgs boson
$h$ is the SM-like Higgs boson observed, and in this scenario 
$c_{\beta - \alpha} $ is constrained close to zero by the current Higgs boson data. 
There have been numerous constraints on 2HDM's. We employ 
\textsc{HiggsSignals}-v2.6.2~\cite{Bechtle:2020uwn} for constraints on the Higgs signal strengths obtained at the LHC 
\cite{Aaboud:2018gay,Aaboud:2018jqu,Aaboud:2018pen,Aad:2020mkp,Sirunyan:2018mvw,Sirunyan:2018hbu,CMS:2019chr,CMS:2019kqw}, 
and \textsc{HiggsBounds}-v5.10.2~\cite{Bechtle:2020pkv} for 
consistency with direct searches at high energy colliders.

\begin{table}[t]
\label{table-yukawa}
\begin{center}
\begin{tabular}{|c||c|c|c||c|c|c||c|c|c|}
\hline
& ~~$\xi^h_u$~~ & ~~$\xi^h_d$~~ & ~~$\xi^h_\ell$~~
& ~~$\xi^H_u$~~ & ~~$\xi^H_d$~~ & ~~$\xi^H_\ell$~~
& ~~$\xi^A_u$~~ & ~~$\xi^A_d$~~ & ~~$\xi^A_\ell$~~ \\ \hline
~~~type-I~~~
& $\frac{c_\alpha}{s_\beta}$ & $\frac{c_\alpha}{s_\beta} $ & $\frac{c_\alpha}{s_\beta} $
& $\frac{s_\alpha}{s_\beta} $ & $\frac{s_\alpha}{s_\beta} $ & $\frac{s_\alpha}{s_\beta} $
& $\frac{1}{t_\beta} $ & $-\frac{1}{t_\beta}$ & $-\frac{1}{t_\beta}$ \\
type-II
& $\frac{c_\alpha}{s_\beta} $ & $-\frac{s_\alpha}{c_\beta}$ & $-\frac{s_\alpha}{c_\beta}$
& $\frac{s_\alpha}{s_\beta}$ & $\frac{c_\alpha}{c_\beta} $ & $\frac{c_\alpha}{c_\beta}$
& $\frac{1}{t_\beta} $ & $t_\beta$ & $t_\beta$ \\
type-III (lepton-specific)
& $\frac{c_\alpha}{s_\beta}$ & $\frac{c_\alpha}{s_\beta}$ & $-\frac{s_\alpha}{c_\beta} $
& $\frac{s_\alpha}{s_\beta} $ & $\frac{s_\alpha}{s_\beta}$ & $\frac{c_\alpha}{c_\beta}$
& $\frac{1}{t_\beta} $ & $-\frac{1}{t_\beta}$ & $t_\beta$ \\
type-IV (flipped)
& $\frac{c_\alpha}{s_\beta} $ & $-\frac{s_\alpha}{c_\beta}$ & $ \frac{c_\alpha}{s_\beta} $
& $\frac{s_\alpha}{s_\beta} $ & $\frac{c_\alpha}{c_\beta}$ & $\frac{s_\alpha}{s_\beta}$
& $\frac{1}{t_\beta} $ & $t_\beta$ & $-\frac{1}{t_\beta}$ \\
\hline
\end{tabular}
\end{center}
\caption{The Yukawa coupling modifiers in the four types of the 2HDM. }\label{table-yukawa}
\end{table}

Conventionally, there are four types of the assignments of the
$Z_2$ parity for the SM fermions, resulting in 2HDM Type I, II, III, and
IV, which differ among themselves in the couplings of Higgs bosons
to fermions. The Yukawa couplings in 2HDM can be parameterized as
\begin{eqnarray}
{\cal L}_{\rm Y} &=&
- \sum_f 
\biggr [
\frac{m_f}{v} \xi^h_f \bar{f} f h + \frac{m_f}{v} \xi^H_f \bar{f} f H
-i \frac{m_f}{v} \xi^A_f \bar{f} \gamma_5 f A
\biggr ] \\ \nonumber
&&
- \biggr[
\frac{\sqrt2 V_{ud} }{v } H^+  \overline{u}
\left(m_u \xi^A_u {P}_L +  m_d \xi^A_d {P}_R\right)d 
+\frac{\sqrt{2} m_\ell}{v}H^+ \xi^A_\ell \overline{\nu_L} \ell_R
+ {\rm H.c.}  \biggr ],
\end{eqnarray}
where the modifiers $\xi^{h,H,A}_f$ are presented in 
Table~\ref{table-yukawa}. 

Since the parameter $\cos(\beta - \alpha)$ is constrained close to zero
by the Higgs boson data, it is instructional to expand the relevant
couplings in terms of $\cos(\beta - \alpha)$. We are considering the
following gluon-fusion process
\[
 pp \to H \to h h \to (b\bar b) (b\bar b) \;.
\]
The relevant couplings include the trilinear coupling $\lambda_{hhH}$,
Yukawa couplings $\lambda^{h,H}_{t,b}$ of $h$ and $H$.
The trilinear coupling $\lambda_{hhH}$ is given by
\begin{equation}
\lambda_{hhH} = \frac{c_{\beta-\alpha} }{s_{2\beta}} \, 
 \left [ s_{2\alpha} (2 m^2_h + M_H^2 ) - \frac{2 m_{12}^2}{ s_{2\beta}}
  (3 s_{2\alpha} - s_{2\beta} ) \right ] 
\end{equation}
The Yukawa couplings can be expanded similarly. For example, in Type II 
they are given by
\begin{eqnarray}
  \lambda^h_t &=& \frac{c_\alpha}{s_\beta} = 
    1 + \frac{c_{\beta-\alpha} } { t_\beta} - 
      \frac{1}{2} c^2_{\beta-\alpha} + {\cal O}(c^3_{\beta-\alpha} )\\
\lambda^h_b &=& - \frac{s_\alpha}{c_\beta} = 
    1 - c_{\beta-\alpha}   t_\beta - 
      \frac{1}{2} c^2_{\beta-\alpha} + {\cal O}(c^3_{\beta-\alpha} )\\
\lambda^H_t &=&  \frac{s_\alpha}{s_\beta} = - \frac{1}{t_\beta} +
    c_{\beta-\alpha}   +
      \frac{c^2_{\beta-\alpha} }{2 t_\beta} + {\cal O}(c^3_{\beta-\alpha})  \\
\lambda^H_b &=&  \frac{c_\alpha}{c_\beta} = t_\beta +
    c_{\beta-\alpha}  - 
      \frac{c^2_{\beta-\alpha} }{2 t_\beta} + {\cal O}(c^3_{\beta-\alpha} ) 
\end{eqnarray}
Other types can be expanded similarly. 
It is straightforward to see that the production cross section via
the resonance CP-even $H$ scales $\cos^2(\beta-\alpha)$. 

Two-Higgs-doublet models are highly constrained by current Higgs signal
strength data from the LHC and direct search bounds on heavy CP-even and
CP-odd scalar bosons, and charged Higgs bosons. We employ the public codes
$\texttt{HiggsBounds-5.10.2}$~\cite{Bechtle:2008jh,Bechtle:2011sb, Bechtle:2012lvg, Bechtle:2013wla, Bechtle:2015pma} and $\texttt{HiggsSignal-2.6.2}$~\cite{Stal:2013hwa,Bechtle:2013xfa,Bechtle:2014ewa,Bechtle:2020uwn} to test the validity of any parameter space points.

\section{Sample Generation and Event Selections}\label{sec:Sample}
\begin{figure}[ht!]
% \begin{center}
\centering
    \begin{subfigure}{0.43\textwidth}
    \centering
    \includegraphics[width=1.\columnwidth]{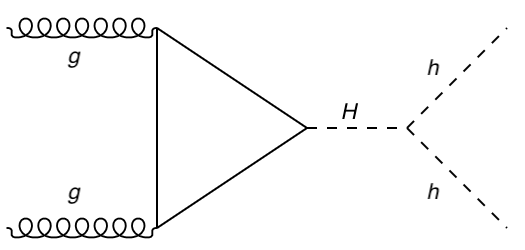}  
    \end{subfigure}
\caption{A Feynman diagram for resonant Higgs boson pair production via gluon fusion in two-Higgs-doublet model.}
\label{fig:feynman_diagram}
% \end{center}
\end{figure}

In this study, the signal is the resonant Higgs boson pair production via gluon fusion in 2HDM, as shown in Fig.\ref{fig:feynman_diagram}. For SM backgrounds, we consider QCD multijet and top-quark pair production($t\bar{t}$) processes to be irreducible SM backgrounds. Other background processes, such as the nonresonant (continuum) SM $hh$
\footnote{We have verified with a parton-level calculation that the resonance peak in the invariant-mass distribution $M_{hh}$ stands tremendously above 
the continuum $hh$ production.}
and electroweak diboson production, are negligible for contributions of event yields~\cite{ATLAS:2022hwc}.

% We consider that the light Higgs boson $h$ ($m_h$ = 125 GeV) pair comes from heavy $CP$-even scalar $H$ with mass $M_H$ = 1000 GeV. The others physical parameters at this benchmark point are $M_A=M_{H^{\pm}}$ = 1000 GeV, $M_{12}^2$ = 400,000 $\text{GeV}^2$, $\tan\beta$ = 5, and $\cos(\beta-\alpha)$ = 0.08. This benchmark point is still allowed under current limits with maximal triple Higgs couplings $\lambda_{hhH}$\cite{Arco:2020ucn} and the largest branching ratio $Br(H\to hh)$ = 0.87 in Type II.

We consider that the light Higgs boson $h$ ($m_h$ = 125 GeV) pair comes from heavy $CP$-even scalar $H$ with mass $M_H$ = 1000 GeV. The others physical parameters at this benchmark point are $M_A=M_{H^{\pm}}$ = 1000 GeV, $M_{12}^2$ = 400,000 $\text{GeV}^2$, $\tan\beta$ = 5, and $\cos(\beta-\alpha)$ = 0.01. This benchmark point is still allowed under current limits\footnote{We verified by $\texttt{HiggsBounds-5.10.2}$ and $\texttt{HiggsSignal-2.6.2}$.} and close to alignment limit in Type II.

\subsection{Monte Carlo Samples}\label{subsec:MC_sample}
The program {\texttt{M{\footnotesize{AD}}G{\footnotesize{RAPH}}5\_{\footnotesize{A}}MC@NLO 2.7.2}}~\cite{Alwall:2014hca} models the signal and background
processes in $pp$ collisions at $\sqrt{s}$ = 14 TeV. The hard-scattering events are passed to {\texttt{P{\footnotesize{YTHIA}} 8.244}}~\cite{Sjostrand:2007gs} to simulate the parton shower and hadronization, using the default settings. According to Ref.\cite{ATLAS:2022hwc}, the {\texttt{NNPDF30\_nlo\_as\_0118}} \cite{NNPDF:2014otw} parton distribution function (PDF) is used for signal in next-to-leading-order calculation and {\texttt{NNPDF23\_lo\_as\_0130\_qed}}\cite{Ball:2012cx} for backgrounds in leading-order calculation.

For signal, the next-to-leading-order two-Higgs-doublet model~\cite{Degrande:2014vpa} is used in the event generation. The input parameters of the benchmark for loop propagators and $s$–channel Higgs boson widths are submitted to {\texttt{M{\footnotesize{AD}}G{\footnotesize{RAPH}}5\_{\footnotesize{A}}MC@NLO 2.7.2}} through parameter cards in the standard setup. The latter are constructed with the public calculator 2HDMC~\cite{Eriksson:2009ws} with $\texttt{HiggsBounds-5.10.2}$~\cite{Bechtle:2008jh,Bechtle:2011sb, Bechtle:2012lvg, Bechtle:2013wla, Bechtle:2015pma} and $\texttt{HiggsSignal-2.6.2}$~\cite{Stal:2013hwa,Bechtle:2013xfa,Bechtle:2014ewa,Bechtle:2020uwn} extensions. Moreover, the decay chain $H\to h h $ is implemented by $\texttt{MadSpin}$\cite{Artoisenet:2012st} and the light Higgs boson $h$ is set to decay 100\% into b$\bar{\mathrm{b}}$ for the moment to obtain the selection efficiency. Later, we use the actual branching ratio
for the event rates.

The $t\bar{t}$ process is simulated at leading order and up to two more jets with the matching scale 20 GeV via MLM prescription\cite{Mangano_2007,Alwall:2007fs}. The other background, the multijet process, is flavor-inclusive. For the multijet process, we require $\texttt{ihtmin}$(inclusive $H_t$ for all partons) is 850 in the $\texttt{run\_card}$ to enhance simulation efficiency. The cross sections of the background samples are normalized based on the event yield given in the ATLAS analysis.

\textsc{Pyjet}~\cite{noel_dawe_2021_4446849,Cacciari:2011ma} and the anti-$k_t$~\cite{Cacciari:2008gp} algorithm with radius parameter $R$ = 1.0 are used to define the boosted jets. After these boosted jets are formed, we require the transverse energy ($E_T$) of the leading large-$R$ jet $>$ 420 GeV and invariant mass of the leading large-$R$ jet $>$ 35 GeV. It is then followed by a trimming procedure~\cite{Krohn:2009th}. The constituents in large-$R$ jets are reclustered into ''subjets" using the $k_T$ algorithm~\cite{Ellis:1993tq} with $R$ = 0.2.  The subjets with less than 5\% of the $p_T$ of the 
large-$R$ jet are then removed. 

After the large-$R$ jets are applied with the trimming procedure, we further setup preselection for each event, following Ref.~\cite{ATLAS:2022hwc}. Each event is required to contain at least two large-$R$ jets with $p_T(J_1)$ $>$ 450 GeV and $p_T(J_2)$ $>$ 250 GeV where $J_1$ and $J_2$ are the leading and subleading jet, respectively. These two large-$R$ jets are also required to have $|\eta(J)|$ $<$ 2 and $M(J)$ $>$ 50 GeV.  After applying the preselection, there remain 600k events from signal and total background for training. In order to generate enough statistics for the analysis, there are 600k events
and 2.3M events from the signal and total background for testing, respectively.

% are in Table.\ref{table:preselection}:
% \begin{table}[h!]
% % \scriptsize
%     \begin{center}
%         \begin{tabular}{c}
%             \hline\hline
%             % transverse momentum of leading jet $p_T(J_1)$ $>$ 450 GeV\\
%             % transverse momentum of subleading jet $p_T(J_2)$ $>$ 150 GeV\\
%             % $|\eta(J_1)|$ $<$ 2 \&  $|\eta(J_2)|$ $<$ 2 \\
%             % invariant mass of leading jet $M(J_1)$ $>$ 50 GeV\\
%             % invariant mass of subleading jet $M(J_2)$ $>$ 50 GeV\\
%             $p_T(J_1)$ $>$ 450 GeV \& $p_T(J_2)$ $>$ 150 GeV\\
%             $|\eta(J_1)|$ $<$ 2 \&  $|\eta(J_2)|$ $<$ 2 \\
%             $M(J_1)$ $>$ 50 GeV \& $M(J_2)$ $>$ 50 GeV\\
%             \hline\hline
%         \end{tabular}
%     \end{center}
% \caption{Preselection table}
% \label{table:preselection}
% \end{table}

% For Test:
% raw: 2000000, 19790000, 12790000
% INFO:root:Before Preselection
% INFO:root:Signal(ppHhh) Length: 1347923
% INFO:root:BKG(ttabr) Length: 1096939
% INFO:root:BKG(ppjjjj) Length: 10269254

% INFO:root:After Preselection
% INFO:root:Signal(ppHhh) Length: 601983
% INFO:root:BKG(ttabr) Length: 346646
% INFO:root:BKG(ppjjjj) Length: 2018501

% For Training:
% INFO:root:Before Preselection
% INFO:root:Signal(ppHhh) Length: 1349109
% INFO:root:BKG(ttabr) Length: 111051
% INFO:root:BKG(ppjjjj) Length: 10269254
% INFO:root:After Preselection
% INFO:root:Signal(ppHhh) Length: 602582
% INFO:root:BKG(ttabr) Length: 35036
% INFO:root:BKG(ppjjjj) Length: 2018501

Furthermore, the Higgs jet is required to satisfy double b-tagging. Jets are declared double $b$-tagged if they have two or more ghosted-associated~\cite{Cacciari:2008gn,Buckley:2015gua} $B$ hadrons. This approach is similar to the subjet b-tagging in ATLAS~\cite{Lin:2018cin,ATLAS:2018sgt}. We do not include the pileup effects in this study. Reference~\cite{Chung:2020ysf} shows that the classification performance is relatively unchanged when neutral particles are not involved. The charged particles are rather insensitive to pileup effects. When setting a precise limit, it would be relatively stable to various experimental effects such as pileup.

% Although it can slightly degrade b-tagging performance\cite{CMS:2017wtu,ATLAS:2017bcq}, when setting a precise limit, it would not change the relative gains presented here.

\subsection{High-level Features}

In order to distinguish the signal from SM backgrounds via Gradient Tree Boosting (BDT), the following fifteen commonly-used high-level features are considered:\\

\quad
1. $M_{JJ}$ : invariant mass of the leading and subleading large-$R$ jets;\\

\quad
2. $M(J_1)$ and $M(J_2)$ : invariant mass of the leading jet and the subleading jet, respectively;\\

\quad
3. $|\Delta\eta(JJ)| \equiv|\eta(J_1) - \eta(J_2)|$;\\

\quad
4. $X_{HH}$: $\sqrt{\left(\frac{M(J_1)-124\text{GeV}}{0.1\times M(J_1)}\right)^2 + \left(\frac{M(J_2)-115\text{GeV}}{0.1\times M(J_2)}\right)^2}$~\cite{ATLAS:2022hwc};\\

\quad
5. $\tau_{21}=\tau_2/\tau_1$ : $n$-subjettiness ratio of the leading jet and the subleading jet~\cite{Thaler:2010tr,Thaler:2011gf};\\
 
\quad
6. $D_2^{(\beta)}=e_3^{(\beta)}/(e_2^{(\beta)})^3$ with $\beta=1,2$ : energy correlation function ratios of the leading jet and the subleading jet~\cite{Larkoski:2014gra};\\

\quad
7. $C_2^{(\beta)}=e_3^{(\beta)}/(e_2^{(\beta)})^2$ with $\beta=1,2$ :  energy correlation function ratios of the leading jet and the subleading jet~\cite{Larkoski:2013eya};\\

\noindent  where $e_i$ is the normalized sum over doublets ($i=2$) or triplets ($i=3$) of constituents inside jets, weighted by the product of the constituent transverse momenta and pairwise angular distances. For this analysis, $\beta$ is considered to be 1 and 2. 

\begin{figure}[ht!]
% \begin{center}
\centering
    \begin{subfigure}{0.43\textwidth}
    \centering
    \includegraphics[width=1.\columnwidth]{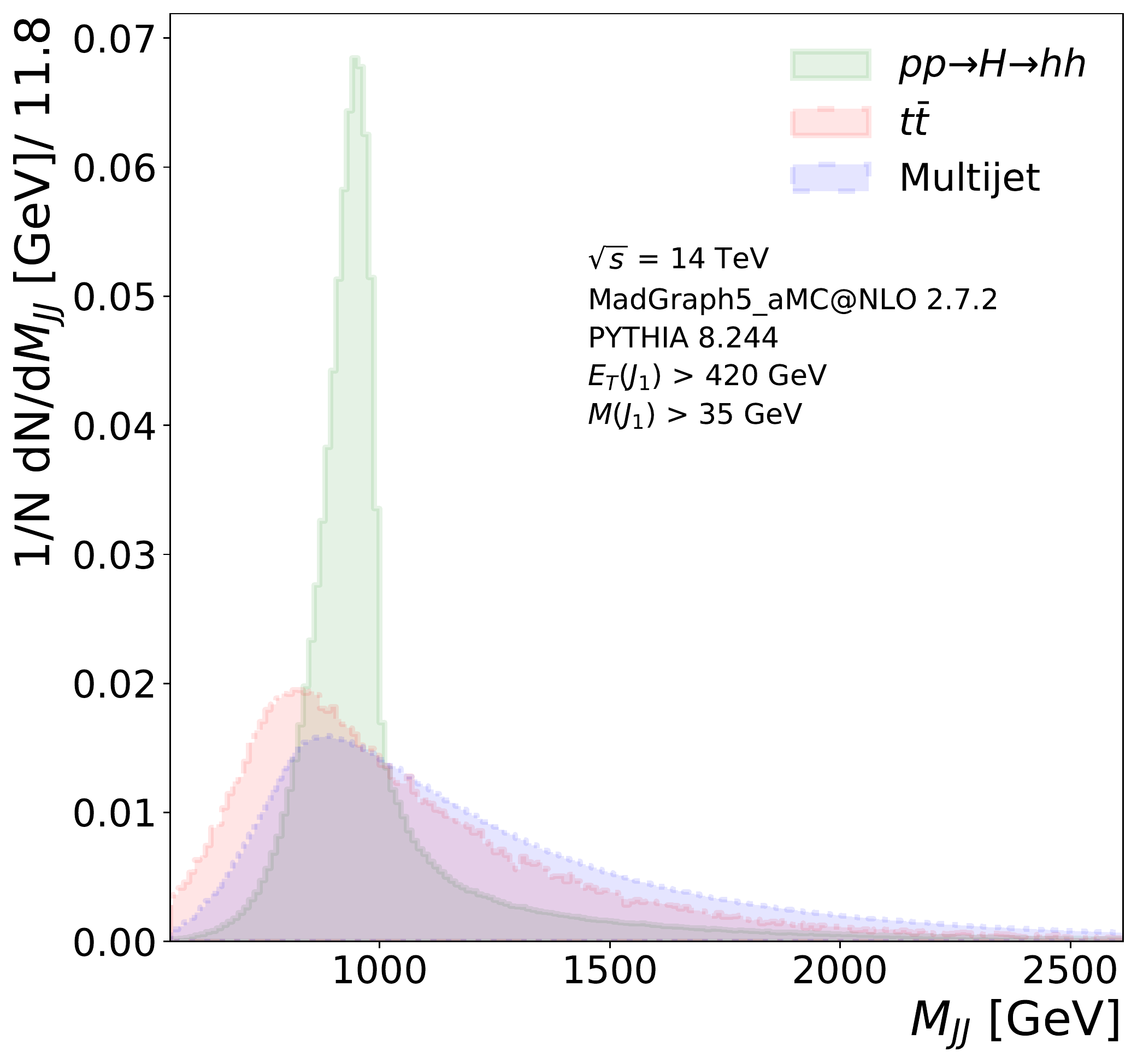}  
    \end{subfigure}
    \begin{subfigure}{0.43\textwidth}
    \centering
    \includegraphics[width=1.\columnwidth]{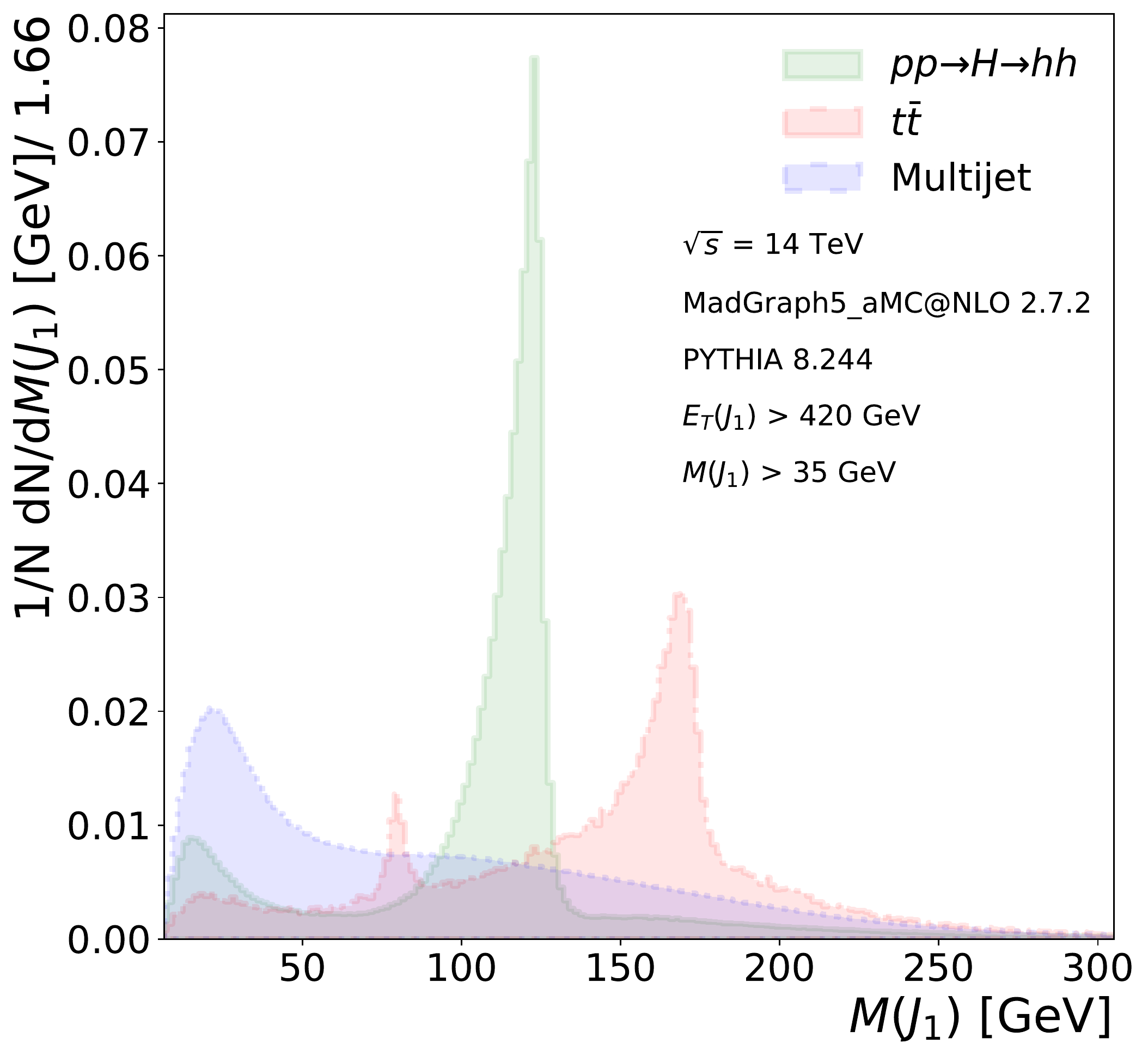}
    \end{subfigure}
    \begin{subfigure}{0.43\textwidth}
    \centering
    \includegraphics[width=1.\columnwidth]{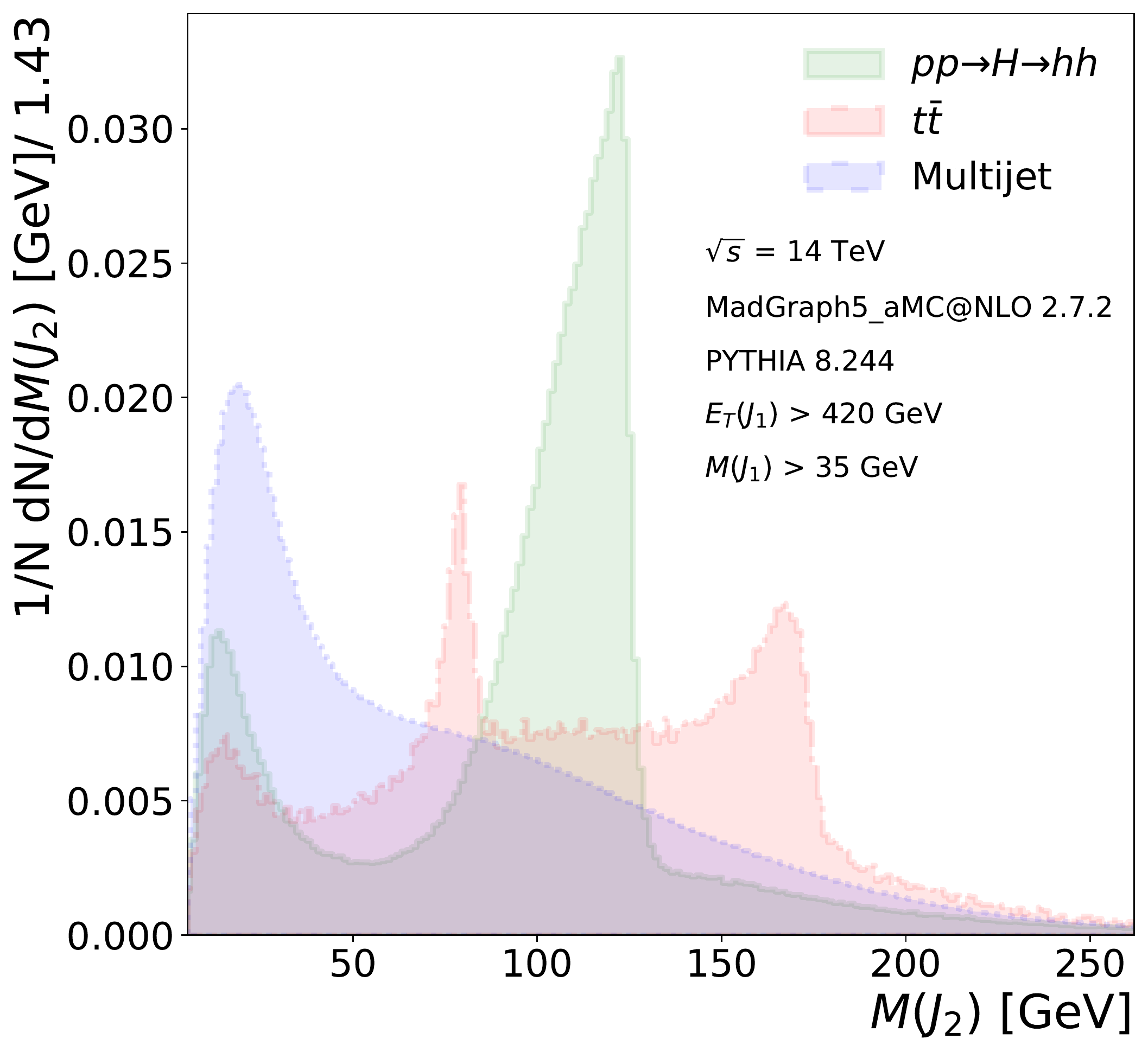}
    \end{subfigure}
    \begin{subfigure}{0.43\textwidth}
    \centering
    \includegraphics[width=1.\columnwidth]{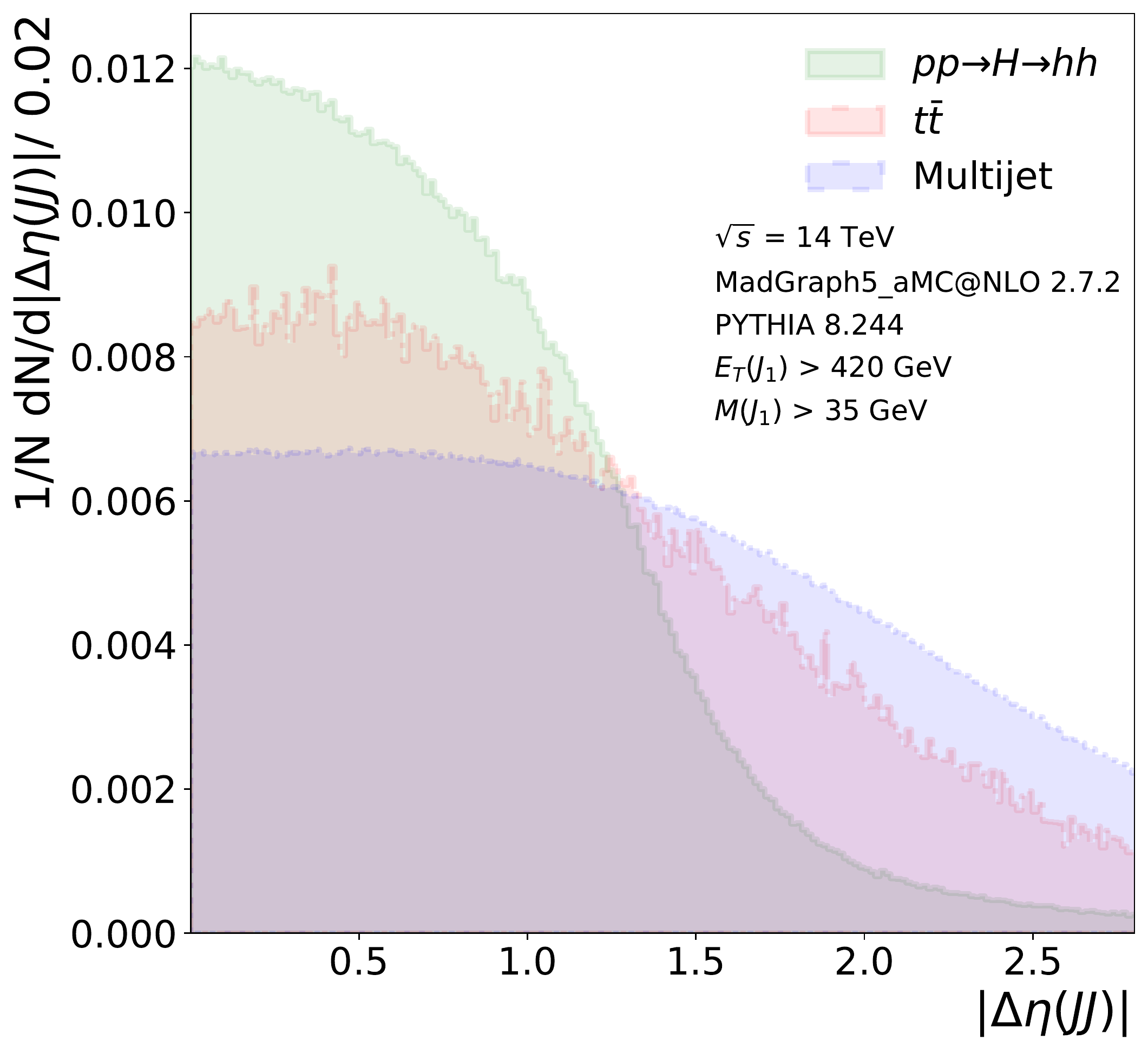}
    \end{subfigure}
    \begin{subfigure}{0.43\textwidth}
    \centering
    \includegraphics[width=1.\columnwidth]{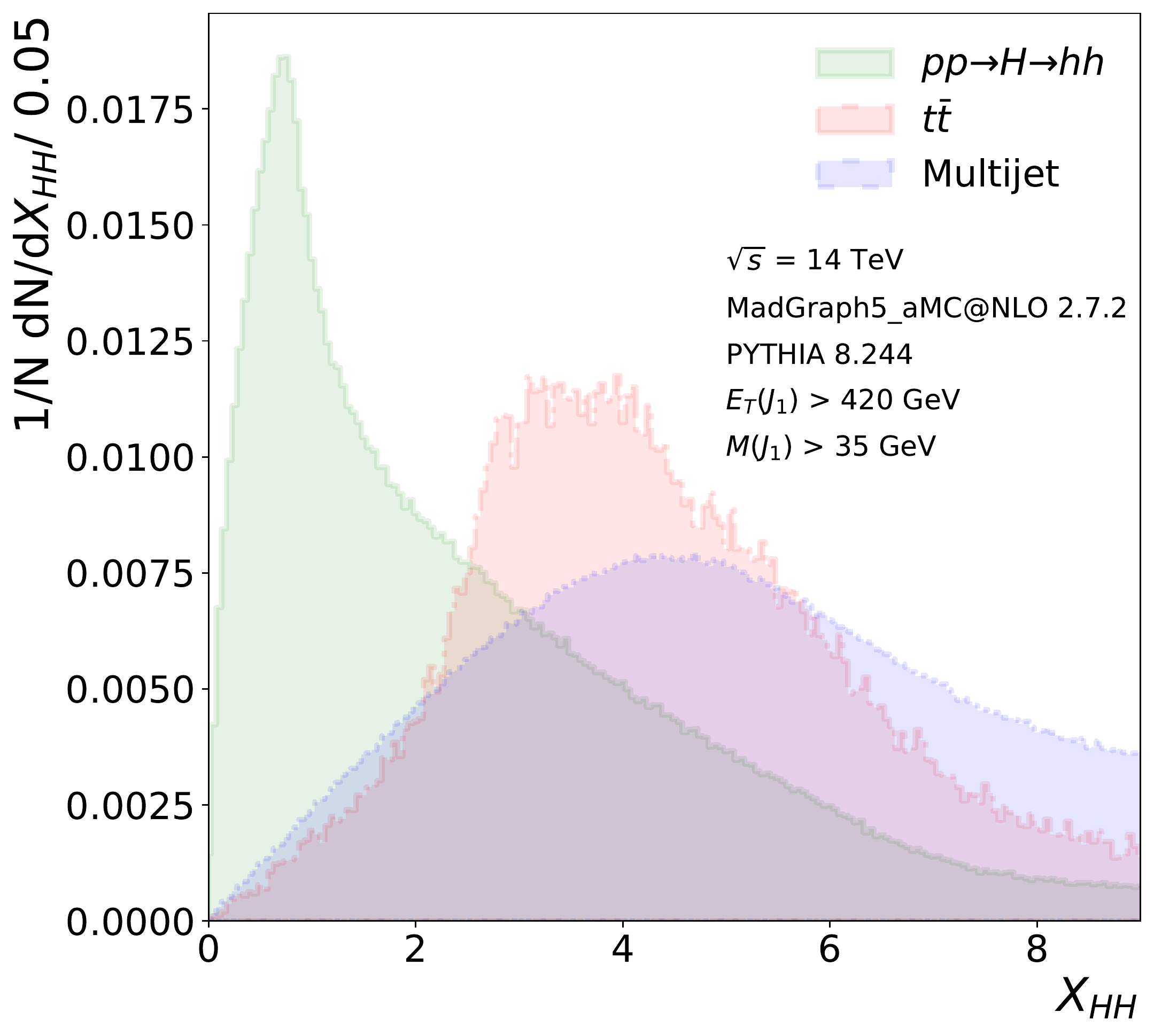}
    \end{subfigure}
\caption{Distributions of the five kinematic variables used in the BDT. In all figures, $E_T(J_1)$ and $M(J_1)$ represent the transverse energy and invariant mass of the leading jet, respectively.}
\label{fig:THDM_kinematic_features}
% \end{center}
\end{figure}

\begin{figure}[ht!]
% \begin{center}
\centering
    \begin{subfigure}{0.43\textwidth}
    \centering
    \includegraphics[width=1.\columnwidth]{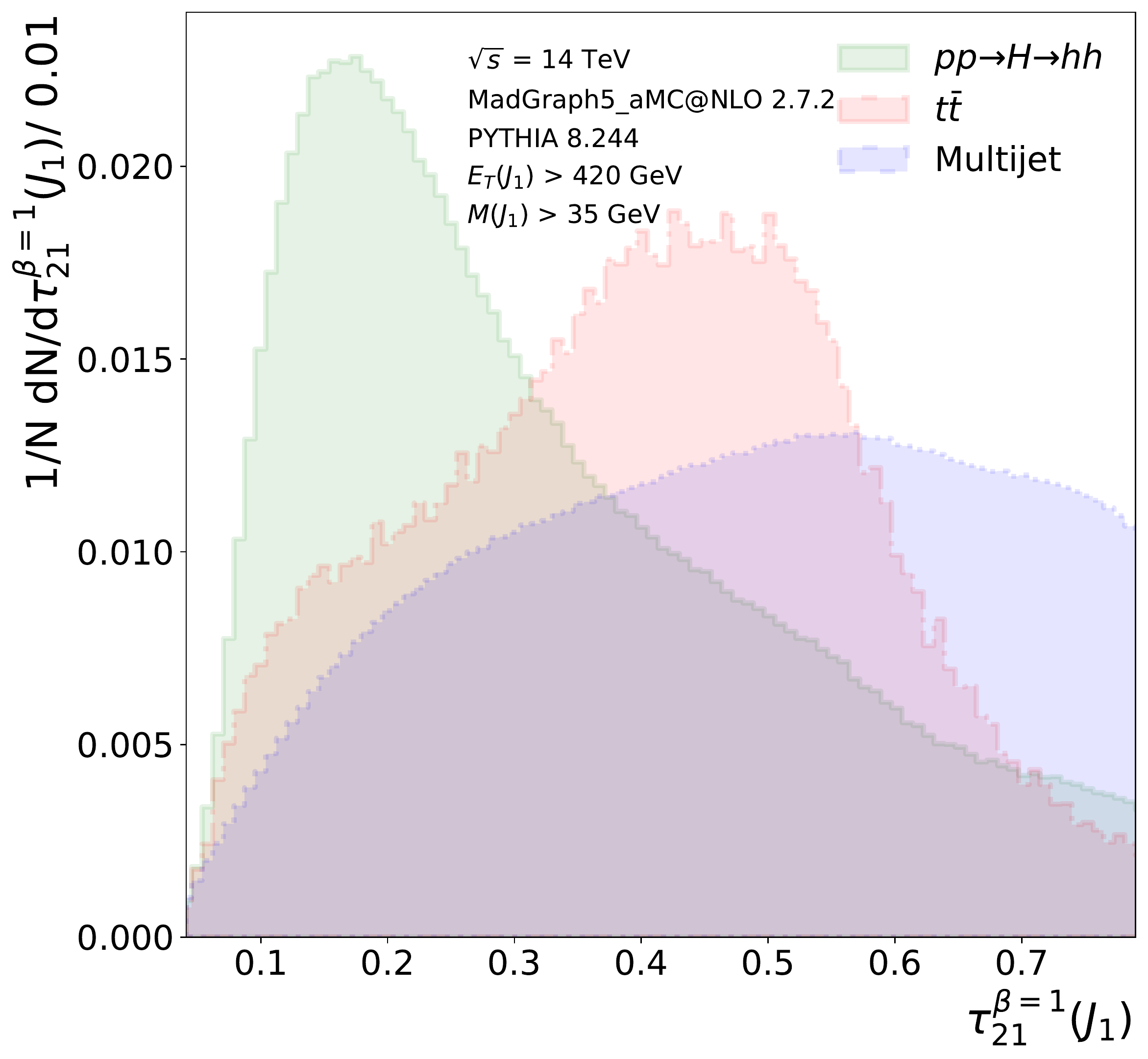}
    \end{subfigure}
    % \begin{subfigure}{0.43\textwidth}
    % \centering
    % \includegraphics[width=1.\columnwidth]{Figures/t212}
    % \end{subfigure}
    \begin{subfigure}{0.43\textwidth}
    \centering
    \includegraphics[width=1.\columnwidth]{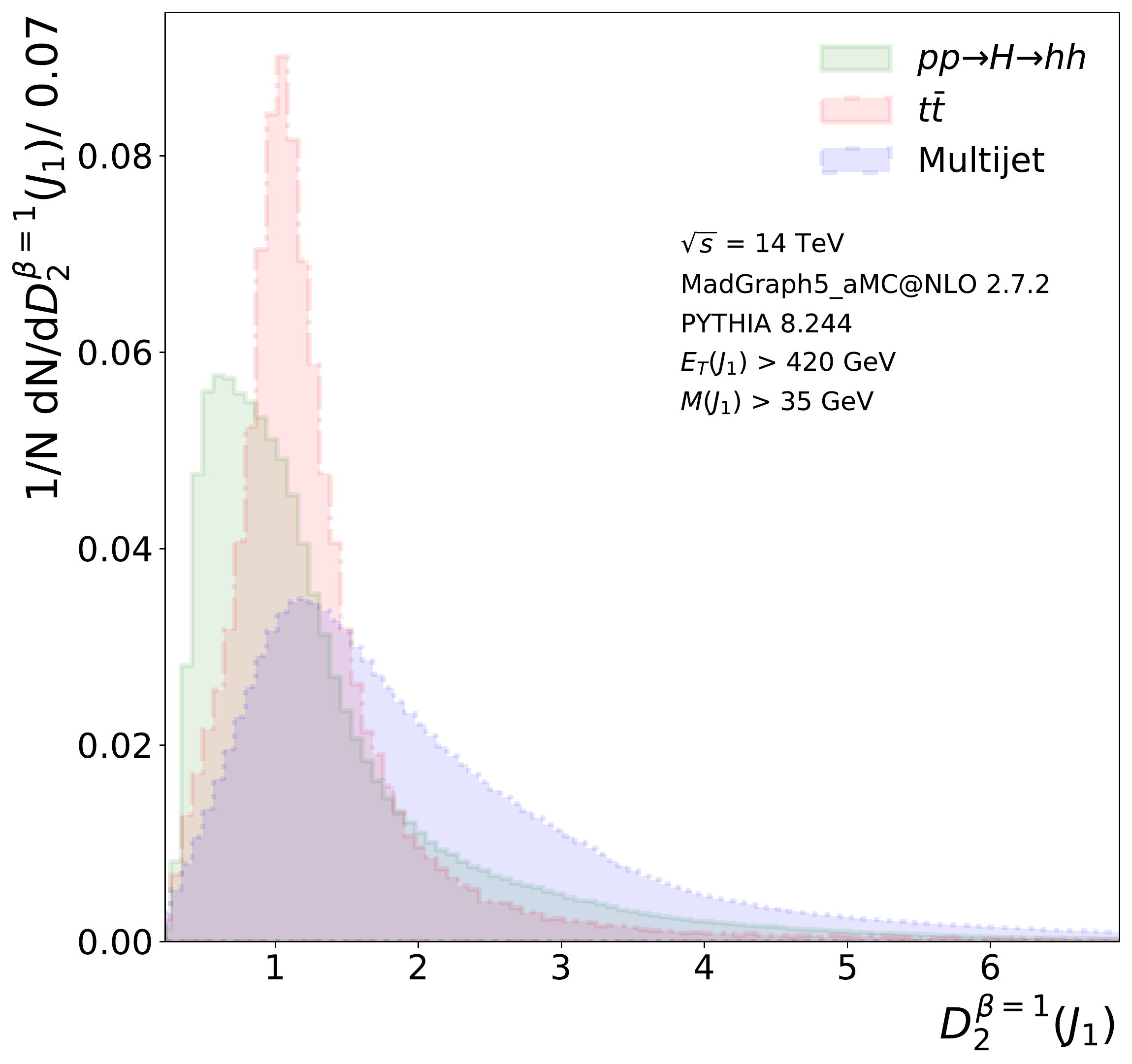}
    \end{subfigure}
    % \begin{subfigure}{0.43\textwidth}
    % \centering
    % \includegraphics[width=1.\columnwidth]{Figures/D212}
    % \end{subfigure}
    \begin{subfigure}{0.43\textwidth}
    \centering
    \includegraphics[width=1.\columnwidth]{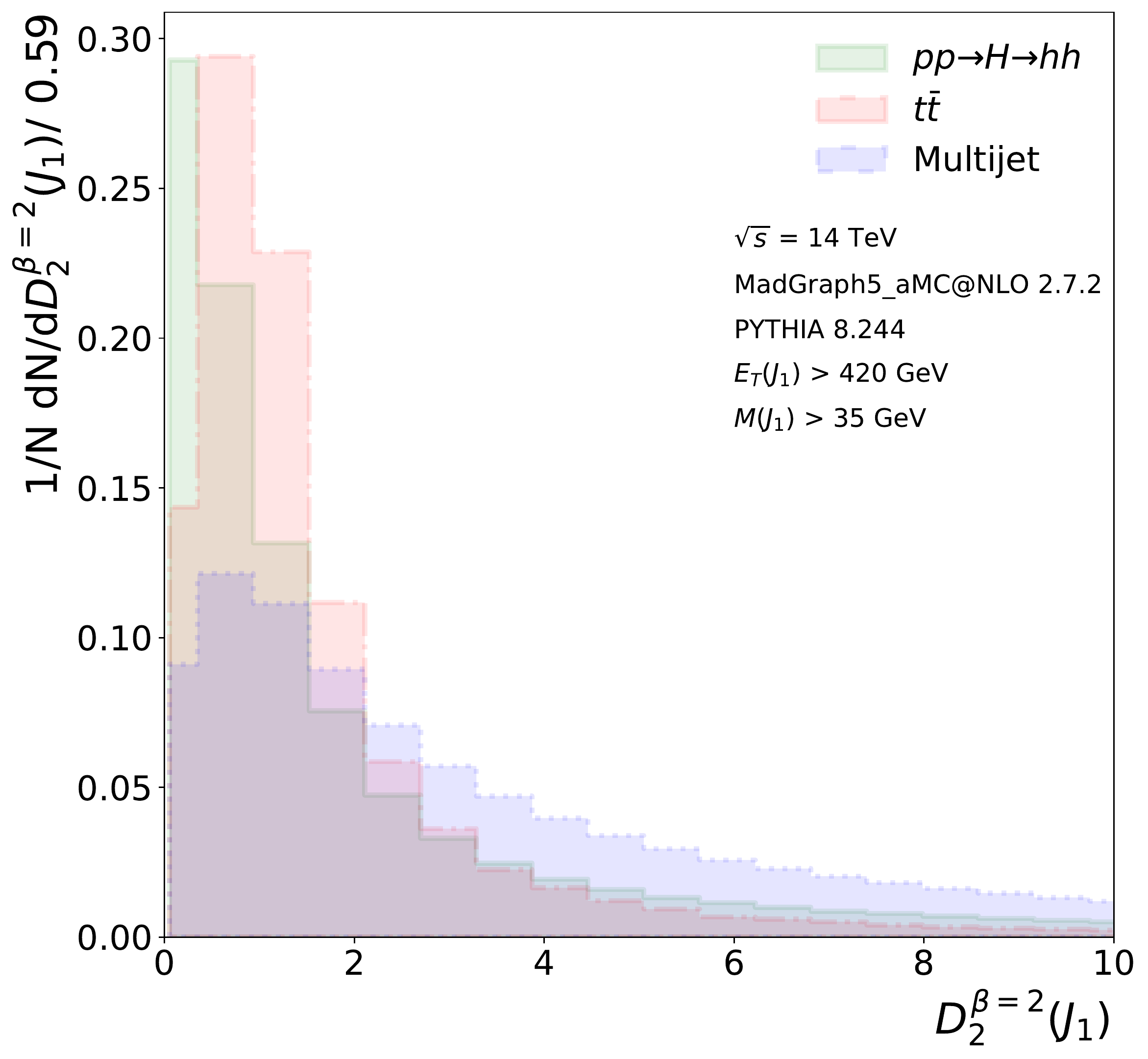}
    \end{subfigure}
    % \begin{subfigure}{0.43\textwidth}
    % \centering
    % \includegraphics[width=1.\columnwidth]{Figures/D222}
    % \end{subfigure}
    \begin{subfigure}{0.43\textwidth}
    \centering
    \includegraphics[width=1.\columnwidth]{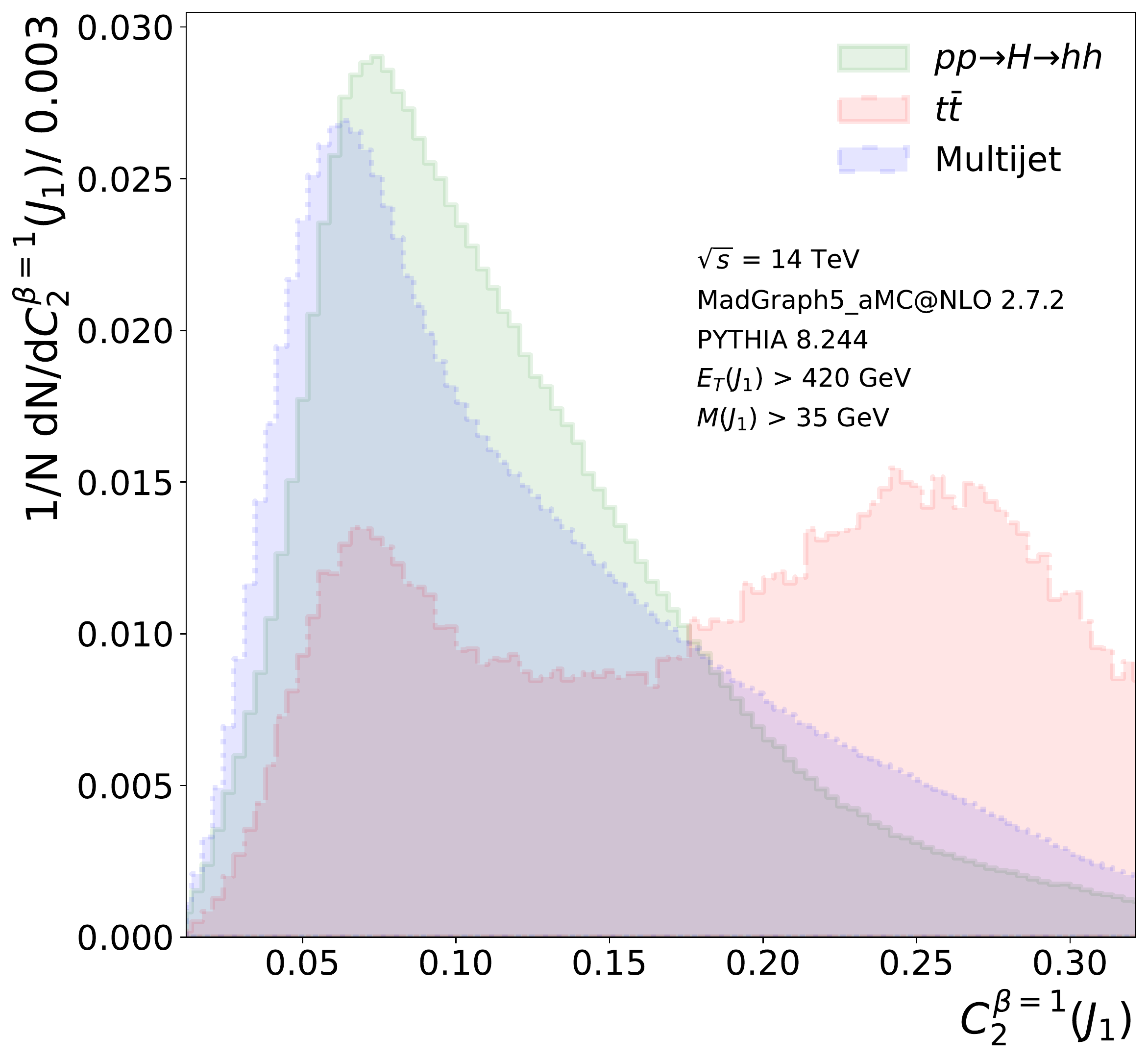}
    \end{subfigure}
    % \begin{subfigure}{0.43\textwidth}
    % \centering
    % \includegraphics[width=1.\columnwidth]{Figures/C212}
    % \end{subfigure}
    \begin{subfigure}{0.43\textwidth}
    \centering
    \includegraphics[width=1.\columnwidth]{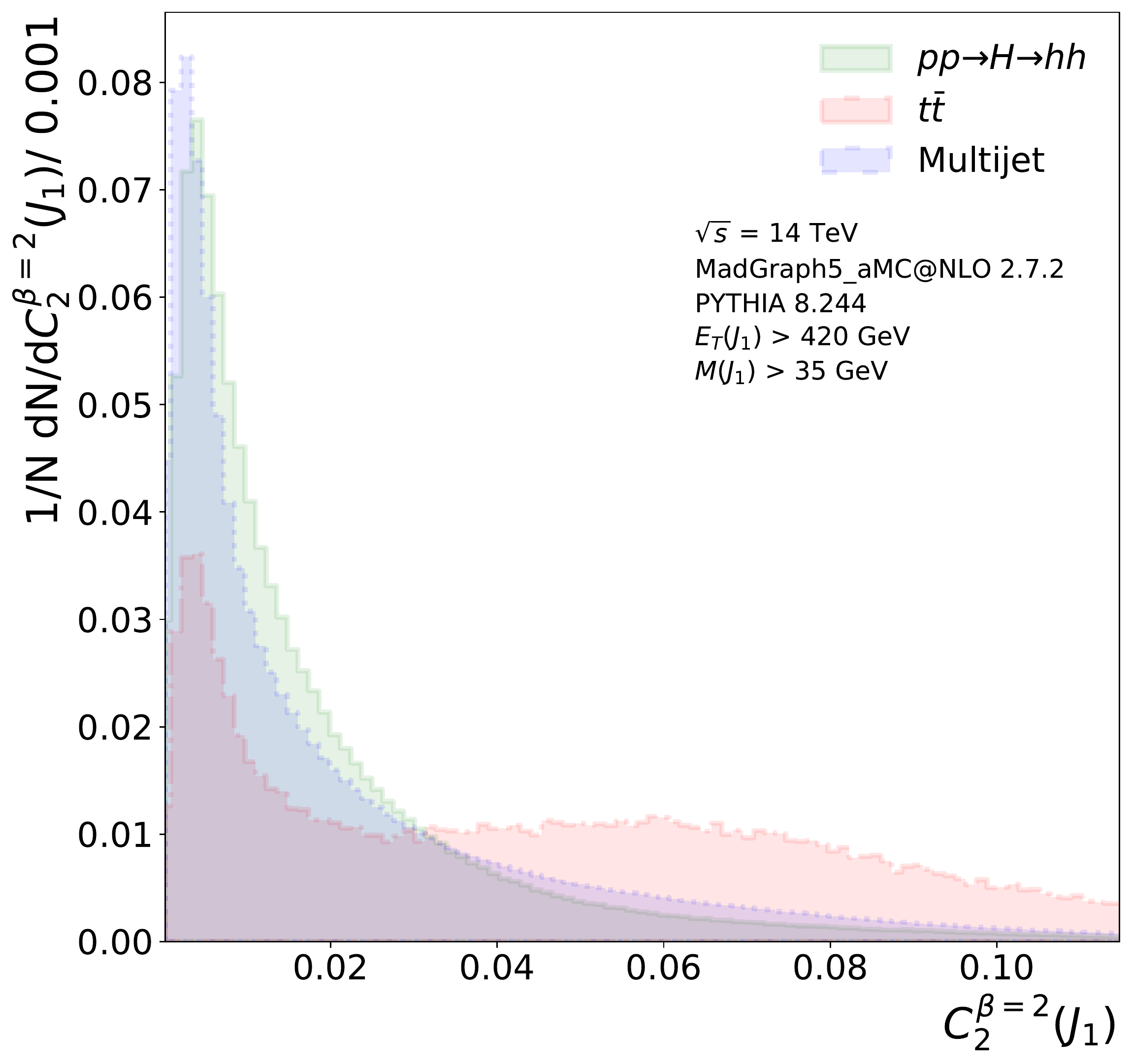}
    \end{subfigure}
    % \begin{subfigure}{0.43\textwidth}
    % \centering
    % \includegraphics[width=1.\columnwidth]{Figures/C222}
    % \end{subfigure}
\caption{Distributions of the five substructures used in the BDT. In all figures, $E_T(J_1)$ and $M(J_1)$ represent the transverse energy and invariant mass of the leading jet, respectively. Here, we only show substructures of the leading jet. The distributions of substructures of the subleading jet are similar to those of the leading jet.}
\label{fig:THDM_substructure_features}
% \end{center}
\end{figure}

The distributions of these variables are shown in Fig.~\ref{fig:THDM_kinematic_features} and Fig.~\ref{fig:THDM_substructure_features}, in which the capability of each observable to discriminate between signal and background is demonstrated. The salient features of these histograms are described in the following. 
The dijet invariant mass distribution peaks near the heavy resonance of 1000 GeV for the signal while it is broad for the background. The resonant signal jets tend to be very central since they are produced through $s$-channel processes. In this case, the $|\Delta\eta(JJ)|$ provides good discrimination power. $X_{HH}$ represents the distance of an event from the di-Higgs peak in the $M(J_1)$-$M(J_2)$ plane. In Fig.~\ref{fig:THDM_kinematic_features}, the peaks of invariant mass of leading jet and subleading jet are around 124 GeV and 115 for signal, respectively. It implies that the signal can be distinguished in the small $X_{HH}$ region. The decay of massive objects into two hard QCD partons produces the two-prong structure, which makes the signal jets result in low $\tau_{21}$,
$D_2^{(\beta)}$ and $C_2^{(\beta)}$.

\subsection{Low-level Features}

\begin{figure}[ht!]
\centering
     \begin{subfigure}{1\textwidth}
        \centering
        \includegraphics[width=1\textwidth]{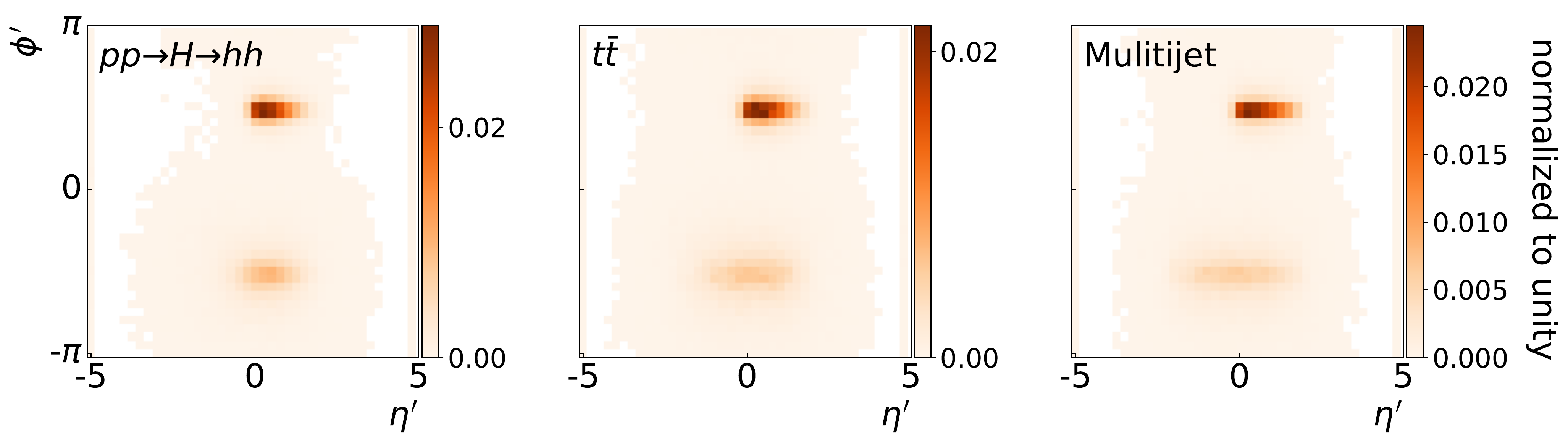}
     \end{subfigure}
     \begin{subfigure}{1\textwidth}
        \centering
        \includegraphics[width=1\textwidth]{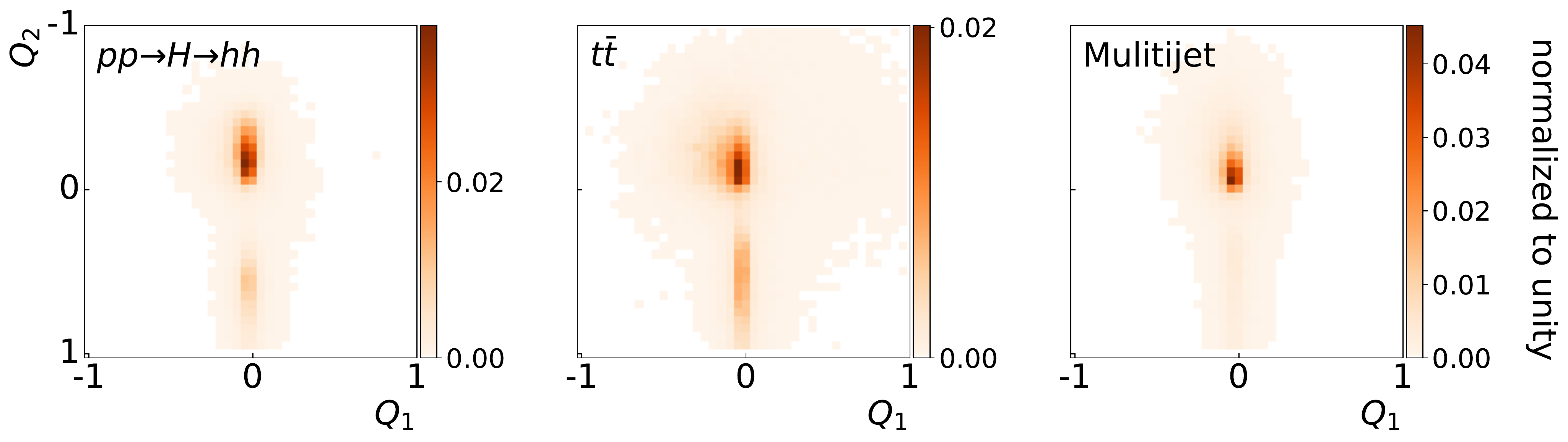}
     \end{subfigure}
     \begin{subfigure}{1\textwidth}
        \centering
        \includegraphics[width=1\textwidth]{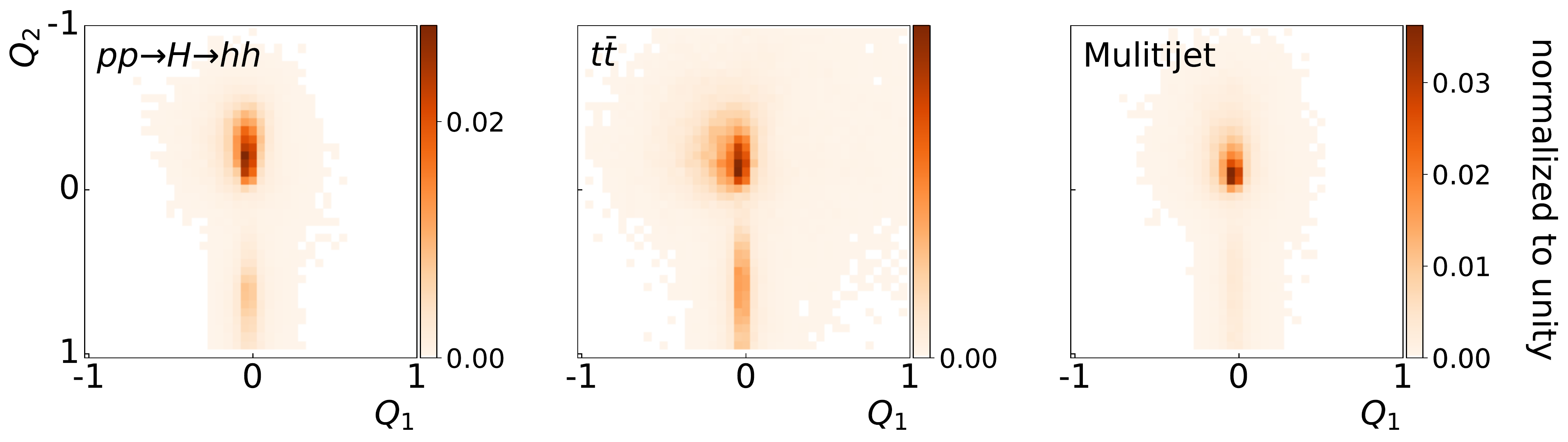}
     \end{subfigure}
\caption{The average of 10000 rotated full-event images (top), leading jet images (middle) and subleading jet images (bottom) in the charged $p_T$ channel. The coordinates $\phi'$ and $\eta'$ denote the new axis after the full-event images are rotated and flipped. $Q_1$ and $Q_2$ denote the new axes after the jet's axis is centralized and rotated. The intensity in each pixel is the sum of the charged particle $p_T$.  The total intensity in each image is normalized to unity. The resolution is 40$\times$40 pixels for each image.}
\label{fig:THDM_lowlevel_features}
\end{figure}

% \newpage
The low-level inputs to the three-stream convolutional neural networks are full-event images, and images of the leading and subleading jets \cite{Cogan:2014oua,deOliveira:2015xxd}. The resolution is 40$\times$40 pixels for both sets of images and jet images are in 1R$\times$1R range\cite{Lin:2018cin, Chung:2020ysf}. The images consist of three channels, analogous to the Red-Green-Blue (RGB) channels of a color image. The pixel intensity for the three channels correspond to the sum of the charged particle $p_T$ , the sum of the neutral particle $p_T$ , and the number of charged particles in a given region of the image. There is no $p_T$ threshold for the contributions to pixel intensity. The full-event image covers effectively the entire $\eta$-$\phi$ cylinder ($|\eta|$ $<$ 5 ). 
Moreover, the full-event images are rotated so that the leading jet is always located at $\phi$ = $\pi$/2. Images are then flipped along the axis defined by $\eta$ = 0 to put the leading jet centroid in the region with positive $\eta$.  The jet images are rotated so that the two subjets are aligned along the same axis. The leading subjet is at the origin and the subleading subjet is directly below the leading subjet. If there is a third-leading subjet, the image will be reflected.
All images are normalized so that the intensities all summed to unity.  After normalization, the pixel intensities are standardized so that their distribution has mean zero and unit variance. These preprocessing procedures significantly improve the stability of the machine learning training\cite{deOliveira:2017pjk}. Figure \ref{fig:THDM_lowlevel_features} shows the average images in the charged $p_T$ channel. The patterns in the charged $p_T$ channel are similar to the other two channels.

\section{Baseline and Classifiers}\label{sec:classifiers}
In this study, there are three selection methods. The first one is conventional cut-based selection following Ref.\cite{ATLAS:2022hwc}. It is called the baseline in this work. The second one is the Gradient Tree Boosting (BDT), 
in which the high-level features will serve as inputs to the BDT. The third 
one, the three-stream convolutional neural networks, which is inspired by Ref.~\cite{Lin:2018cin, Chung:2020ysf} will demonstrate the discriminating power in this work.

\subsection{The Cut-based Method}
We treat the similar boosted channel analysis in Ref.\cite{ATLAS:2022hwc} as the baseline in this study. After preselection which is described in Sec.\ref{subsec:MC_sample}, we further apply $|\Delta\eta(JJ)|$ $<$ 1.3 and $X_{HH}$ $<$ 1.6 to reduce SM backgrounds. Then we require $M_{JJ}$ should be in the heavy resonance mass window: 900 GeV $<$ $M_{JJ}$ $<$ 1100 GeV. After this mass window selection for the heavy resonance, the signal efficiency contains roughly 90\% before applying this criteria. At last, we require the leading and subleading jets should be the Higgs jets.

\subsection{The Boosted Decision Tree}
In this study, the BDT uses Gradient Tree Boosting. It has a fixed number of estimators (2000)  with maximum depth 5. The minimum number of samples is fixed at 25\% as required to split an internal node and 5\% as required to be at a leaf node. The deviance of the loss function is set with the learning rate 0.005. This BDT model is trained on fifteen high-level features of the jet using the {\texttt{scikit-learn}} library \cite{scikit-learn}.

\subsection{The Three-stream Convolutional Neural Networks (3CNN)}
\begin{figure}[t!]
% \begin{center}
\centering
    \begin{subfigure}{0.9\textwidth}
    \centering
    \includegraphics[width=1.\columnwidth]{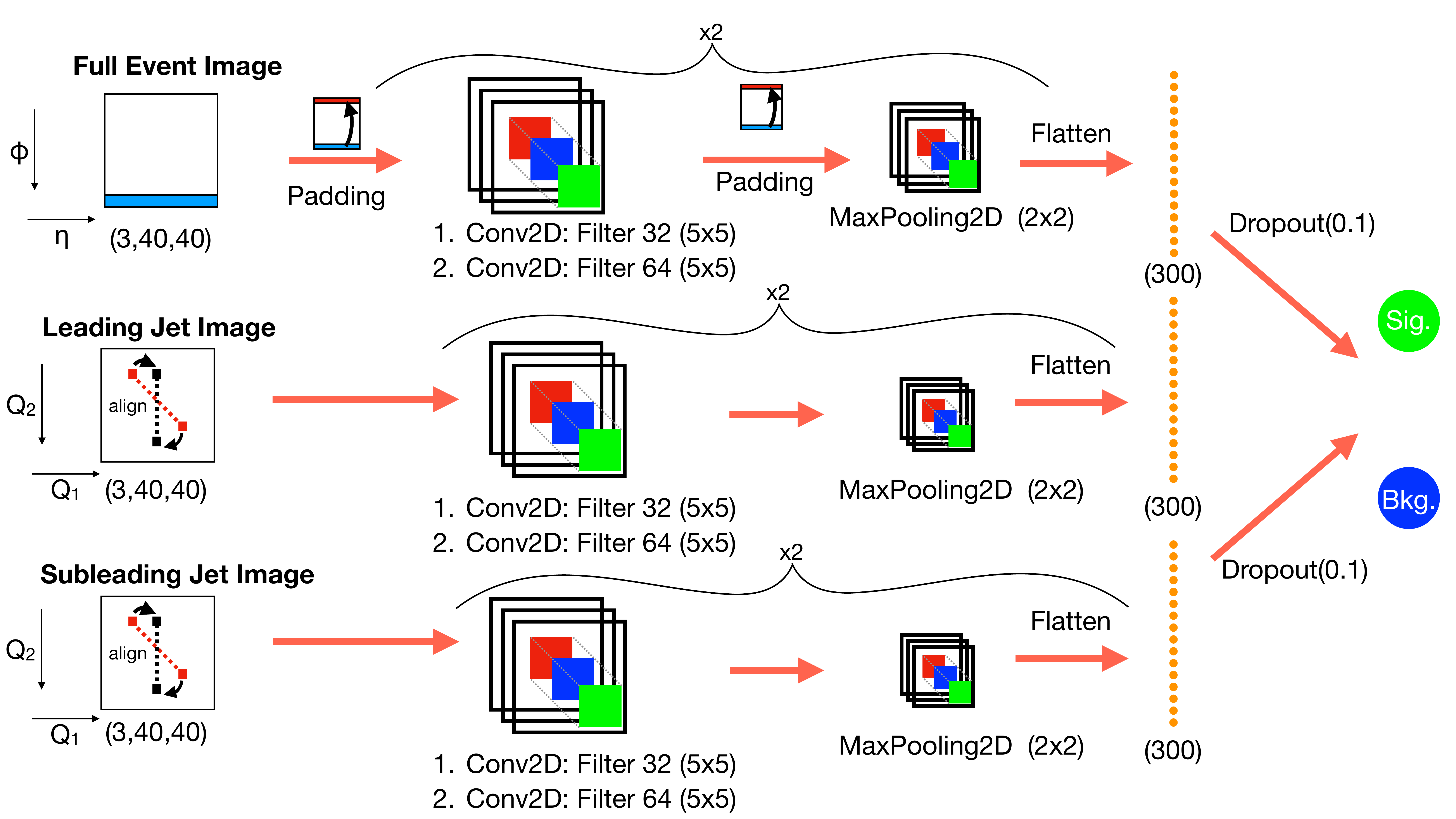}  
    \end{subfigure}
\caption{Architecture of the 3CNN, based on Ref.\cite{Lin:2018cin, Chung:2020ysf}. The first stream (top) is used to process full-event images. The second stream (middle) uses the information
from the leading jet. The third stream (bottom) uses the information
from the subleading jet.}
\label{fig:3CNN_architecture}
% \end{center}
\end{figure}

The 3CNN in this study is based on Ref.\cite{Lin:2018cin, Chung:2020ysf}. One stream of the 3CNN is dedicated to global full-event information. The other two streams are dedicated to processing local information in the leading jet and subleading jet. In addition, there are two outputs in the last layer for disentangling the signal and SM backgrounds. The three-stream architecture is shown schematically in Fig.~\ref{fig:3CNN_architecture}.

Details of the 3CNN are as follows. The convolution filter is 5$\times$5 in three streams, the maximum pooling layers are 2$\times$2, and the stride length is 1. Rectified linear unit (ReLU) activation functions are used for all intermediate layers of the neural network (NN). The first convolution layer in each stream has 32 filters and the second convolution layer in each stream has 64 filters. There are 300 neurons for the dense layer at the end of each stream. The three dense layers from each stream are fully connected to two output neurons with the softmax activation function $e^{x_i}/\sum_{i=1}^4 e^{x_i}$, which is the multidimensional generalization of the sigmoid.  The AdaDelta optimizer~\cite{DBLP:journals/corr/abs-1212-5701} is used to select the network weights. Between the last dense layer and output layer, Dropout~\cite{JMLR:v15:srivastava14a} regularization is added to reduce overfitting with the dropout rate = 0.1. The categorical cross entropy loss function is optimized in the neural network training. For effectively utilizing the full information of the detector in the $\phi$ direction, a padding method is used to take the information in the bottom four rows of the input images and append them onto the top of the image. The {\texttt{Keras-2.4.0}} library is used to train a 3CNN model with the {\texttt{T{\footnotesize{ENSORFLOW}}-2.4.0-rc3}} \cite{tensorflow2015-whitepaper} backend, on a {\texttt{NVIDIA RTX A6000 48 GB}}.

% \section{Analysis and results at 14 TeV HL-LHC}\label{sec:scanning}
\section{Results and Sensitivity Reach in 2HDM}\label{sec:Sensitivity}

The selection for signal and SM backgrounds are shown in Table.\ref{table:total_selection}. Here we set light Higgs boson mass $m_h$ = 125 GeV, heavy $CP$-even scalar mass $M_H$ = 1000 GeV, $M_A=M_{H^{\pm}}$ = 1000 GeV, $M_{12}^2$ = 400,000 $\text{GeV}^2$, $\tan\beta$ = 5, and $\cos(\beta-\alpha)$ = 0.01 in Type II for the benchmark point. We list two major SM backgrounds in this table: $t\bar{t}$ and multijet. The multijet 
background is the dominant one between them before applying the selection. The notation $\textbf{preselection}$ in Table.\ref{table:total_selection} is the cut flow that was described in the Sec.\ref{subsec:MC_sample}. We apply extra B-hadron tagging efficiency = 0.77~\cite{ATLAS:2022hwc} for the Higgs jet, requiring double b-tagging via ghosted-associated method, to estimate the event yield. Moreover, in order to compare the background-discriminating power 
among Baseline, BDT and 3CNN, we choose the BDT score cut and 3CNN score cut to make the number of signal events be close to that in the baseline analysis. 

From Table.\ref{table:total_selection}, we find that the BDT analysis outperforms the baseline analysis based on the cut-based method. The number of signal events in this benchmark point is around 28 while the total background
is around 1390 events in the baseline analysis. On the other hand, in the BDT analysis the efficiency of the signal is about the same, while the 
background rejection power improves by a factor of 10 over the cut-based method.
It turns out that the number of signal event in this benchmark point is around 25 and the total background is around 140 events in the BDT analysis. After we introduce the 3CNN analysis, we still can maintain 25 signal events but the total background is reduced to 56 events. 

\begin{table}[h!]
\scriptsize
\begin{center}
\begin{tabular}{cccccc}
\hline\hline
\multicolumn{6}{c}{\textbf{Selection Flow Table}}\\
\hline
\multicolumn{2}{c}{} &\textbf{$pp\to H\to h h \to b\bar{b}b\bar{b}$ (Type II)}&\textbf{$t\bar{t}$}&\textbf{Mulitijet}& \textbf{Total Backgrounds}\\
\hline
\multicolumn{2}{c}{\textbf{preselection}} & $8.02\times10^{1}$ & $9.23\times10^{5}$ & $2.76\times10^{7}$ & $2.86\times10^{7}$\\
\multicolumn{2}{c}{\textbf{900 GeV $<$ $M_{JJ}$ $<$ 1100 GeV}} & $5.29\times10^{1}$ & $2.77\times10^{5}$ & $6.92\times10^{6}$ & $7.20\times10^{6}$\\
\multicolumn{2}{c}{\textbf{2 Higgs jets}} & $4.74\times10^{1}$ & $1.05\times10^{3}$ & $2.34\times10^{4}$ & $2.45\times10^{4}$\\
\hline\hline
\multirow{2}{*}{\textbf{Baseline}} &\textbf{$|\Delta\eta(JJ)|$ $<$ 1.3} & $4.68\times10^{1}$ & $9.99\times10^{2}$ & $2.18\times10^{4}$ & $2.28\times10^{4}$\\
&\textbf{$X_{HH}$ $<$ 1.6}& $2.82\times10^{1}$ & $2.13\times10^{1}$ & $1.37\times10^{3}$ & $1.39\times10^{3}$\\
\hline\hline
\multicolumn{2}{c}{\textbf{BDT score $>$ 0.964}} &  $2.56\times10^{1}$ & $5.33$ & $1.37\times10^{2}$ & $1.42\times10^{2}$ \\
\hline\hline
\multicolumn{2}{c}{\textbf{3CNN score $>$ 0.99}} &  $2.56\times10^{1}$ & $2.93\times10^{1}$ & $2.74\times10^{1}$ & $5.67\times10^{1}$ \\
\hline\hline
\end{tabular}
\end{center}
\caption{Table showing the cut flow and event yield for the signal process
$pp \to H \to h h \to b \bar b b \bar b$ and the backgrounds at $\sqrt{s}$ = 14 TeV with an integrated luminosity $\mathcal{L}$ = 3000 $fb^{-1}$. The signal is Type II of 2HDMs. 
The B-hadrons tagging efficiency = 0.77~\cite{ATLAS:2022hwc} is applied to calculate the event yield. The preselection are 
described in the main text.}
\label{table:total_selection}
\end{table}

\begin{figure}[h!]
% \begin{center}
\centering
    \begin{subfigure}{0.49\textwidth}
    \centering
    \includegraphics[width=1.\columnwidth]{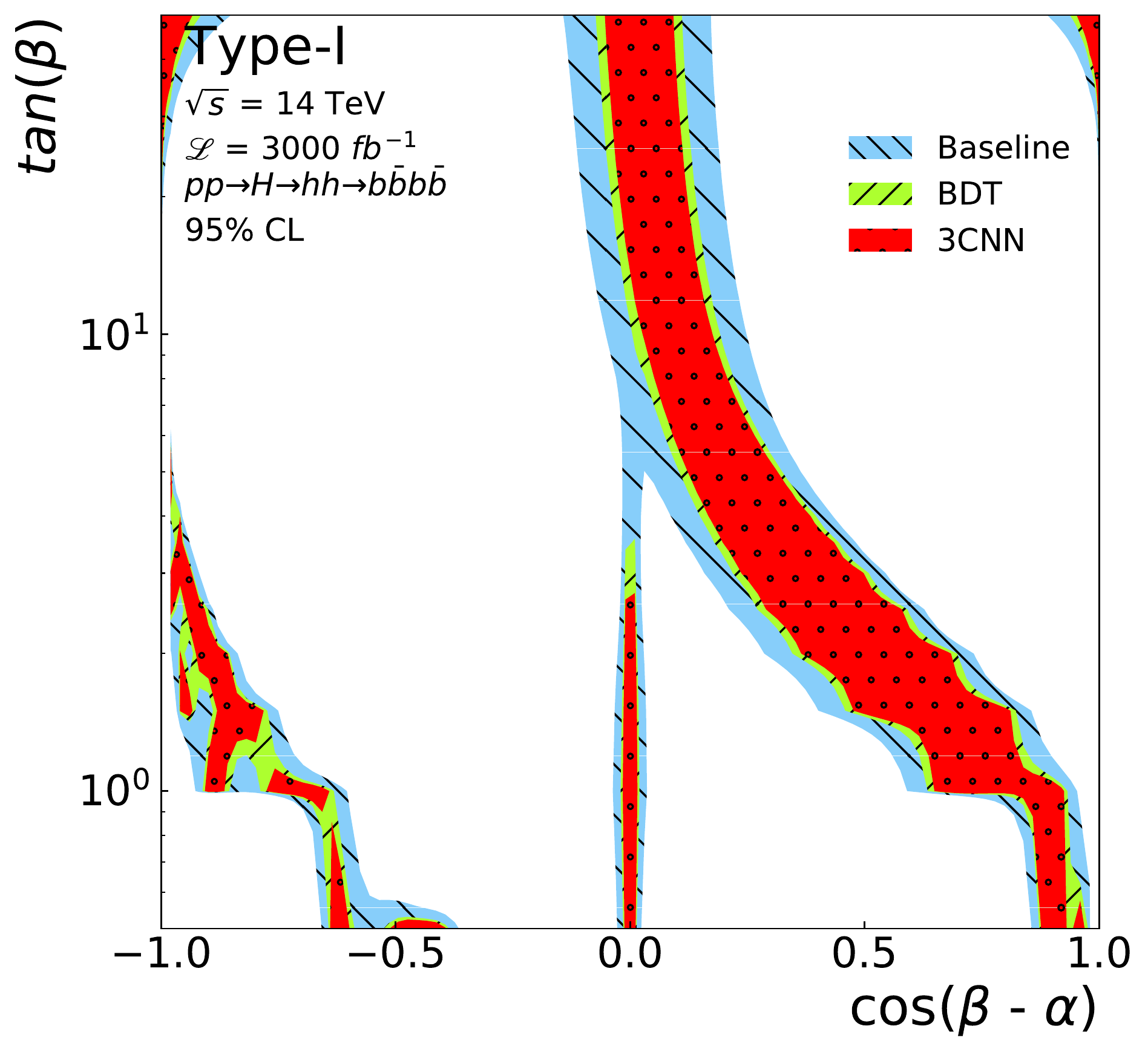}  
    \end{subfigure}
    \begin{subfigure}{0.49\textwidth}
    \centering
    \includegraphics[width=1.\columnwidth]{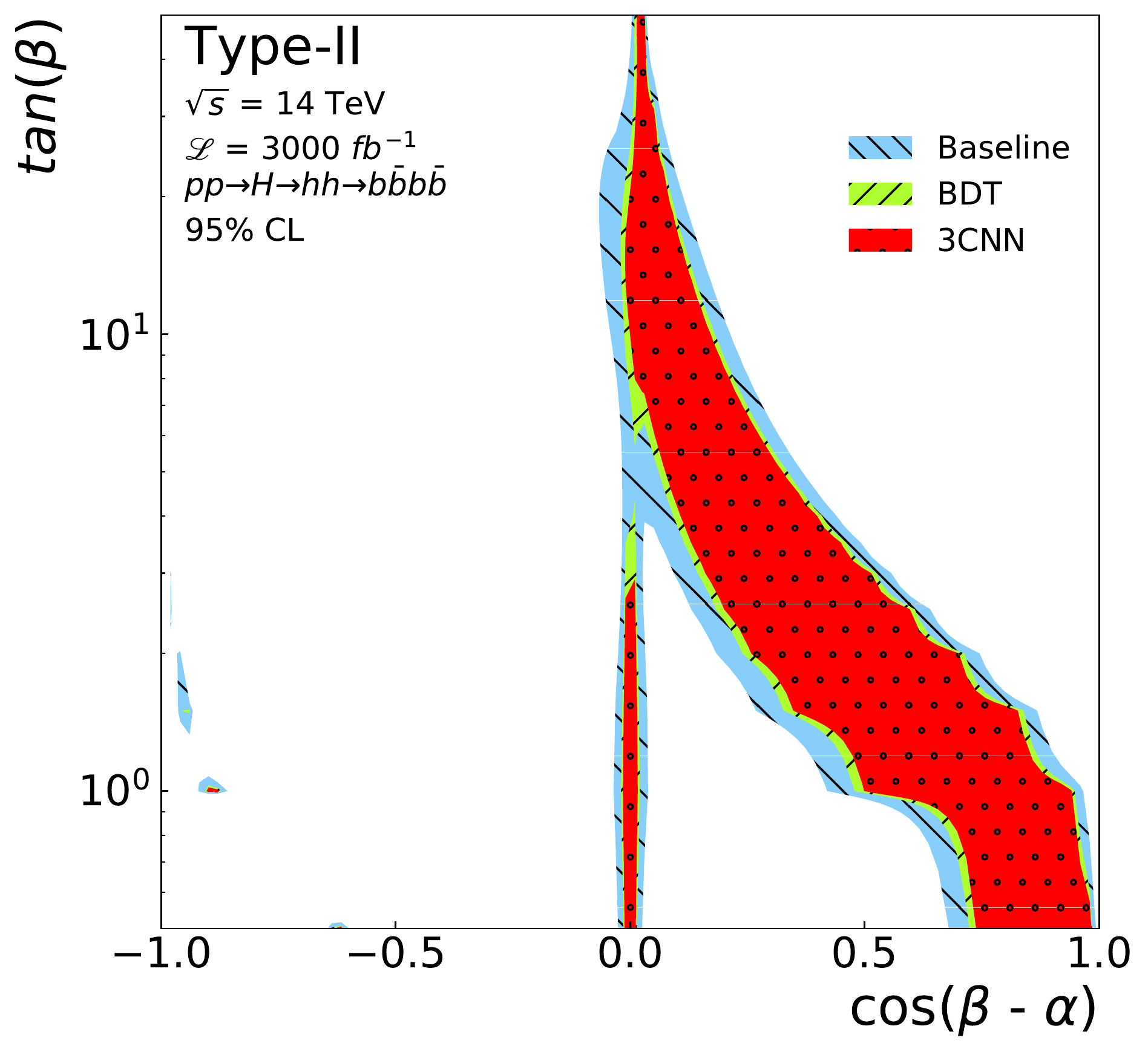}  
    \end{subfigure}
    \begin{subfigure}{0.49\textwidth}
    \centering
    \includegraphics[width=1.\columnwidth]{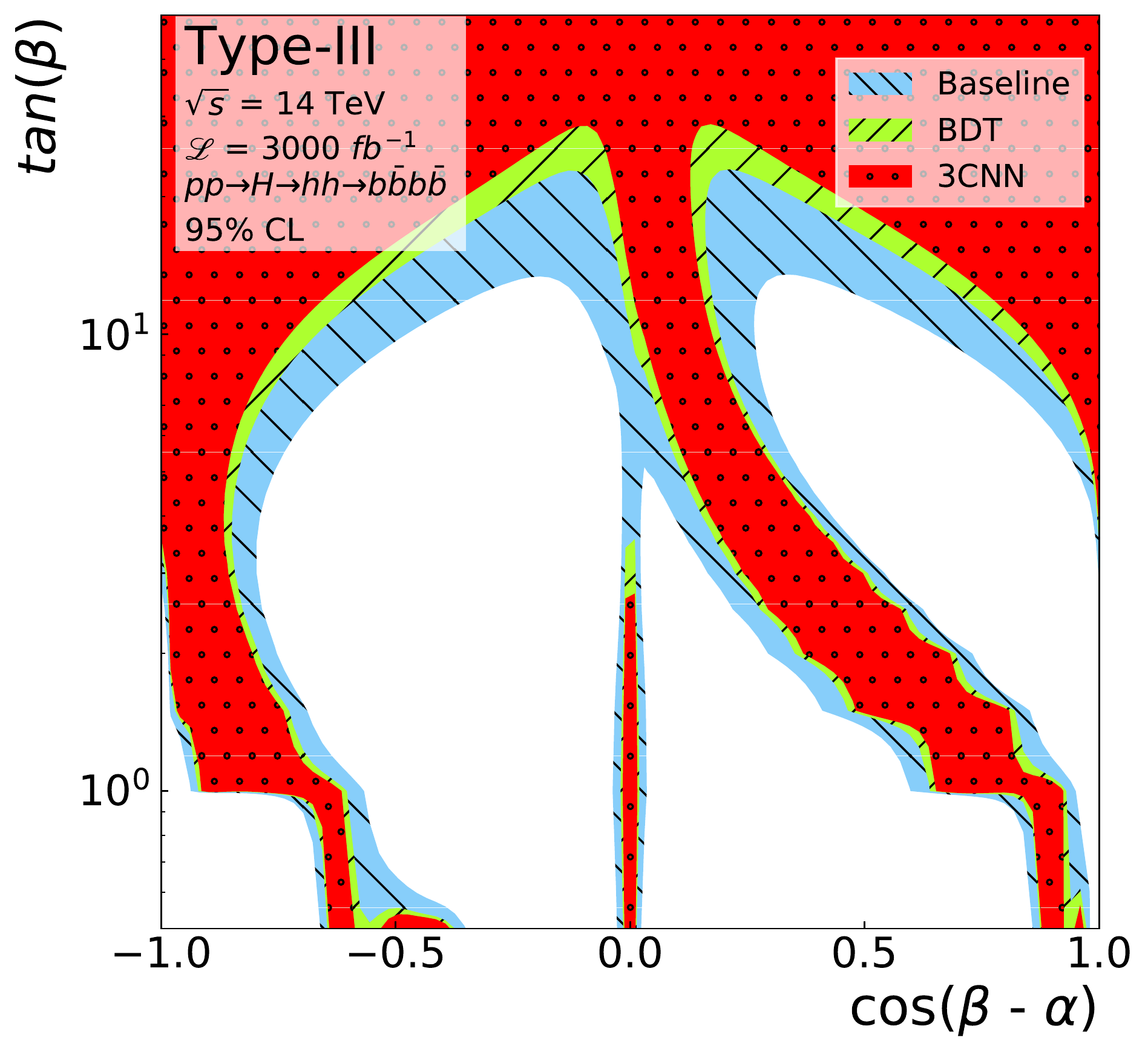}  
    \end{subfigure}
    \begin{subfigure}{0.49\textwidth}
    \centering
    \includegraphics[width=1.\columnwidth]{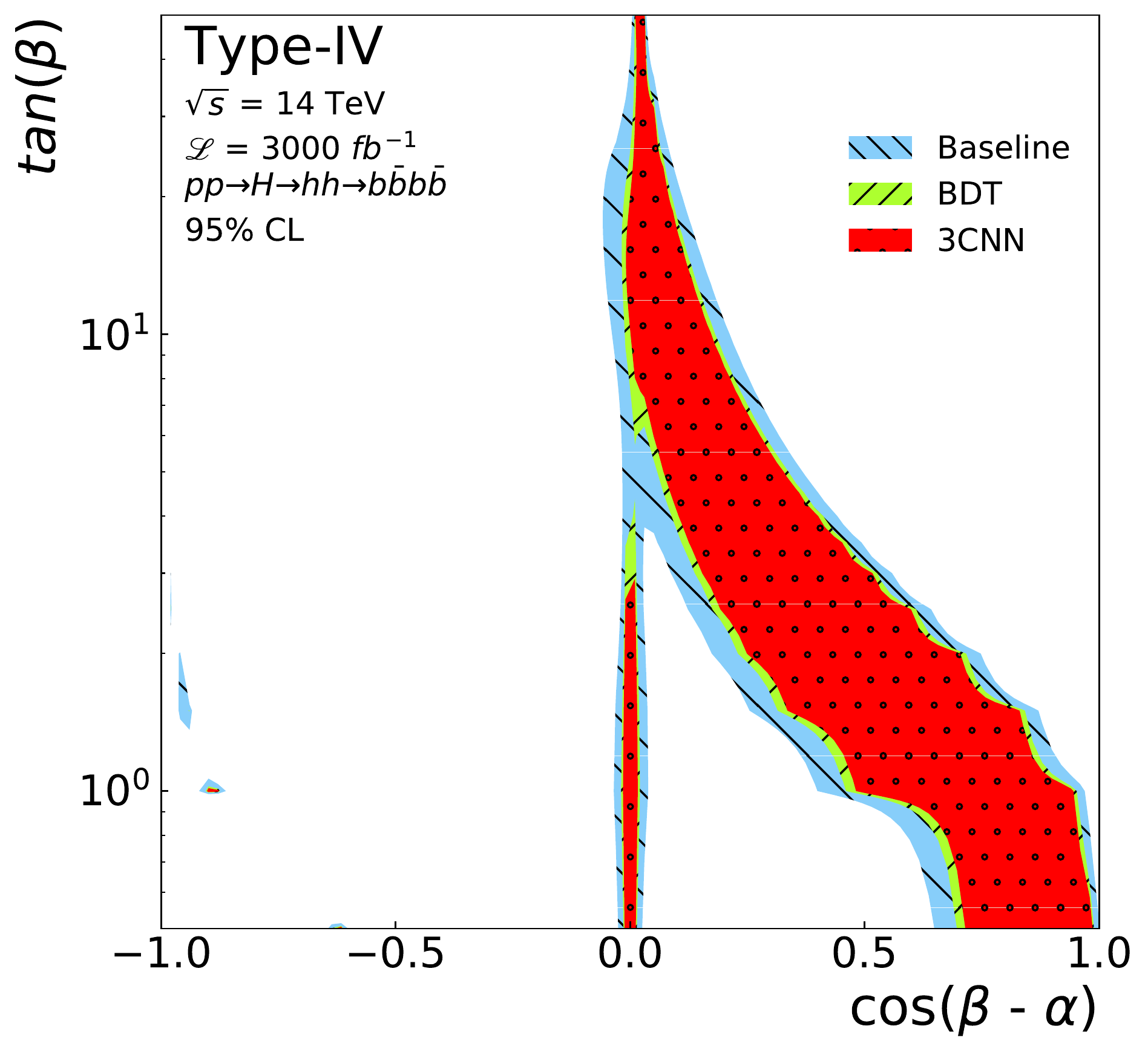}  
    \end{subfigure}
\caption{Allowed regions in all four types of 2HDM for $M_A=M_{H^{\pm}}$ = 1000 GeV, $M_{12}^2$ = 400,000 $\text{GeV}^2$ in the ($\cos(\beta-\alpha)$, $\tan\beta$) plane if no excess is seen above the SM background at the HL-LHC.
The blue, green and red regions are the allowed region based on baseline, BDT and 3CNN analysis, respectively. Allowed region is the area with significance $\le$ 2, significance is $\sqrt{2[(s+b)ln(1+s/b)-s]}$, where $s$ is the number of signal events and $b$ is the number of background events.}
\label{fig:allowed_tb_cba}
% \end{center}
\end{figure}

Results of these three analysis at $\sqrt{s}$ = 14 TeV with an integrated luminosity $\mathcal{L}$ = 3000 $fb^{-1}$ are interpreted in the
parameter space ($\cos(\beta-\alpha)$, $\tan\beta$) in Fig.\ref{fig:allowed_tb_cba}. 
We fix $M_A=M_{H^{\pm}}$ = 1000 GeV, 
and $M_{12}^2$ = 400,000 $\text{GeV}^2$ to find the allowed
region at 95\% CL in the ($\cos(\beta-\alpha)$, $\tan\beta$) plane. 
Note that the colored regions are those with the significance
$z \equiv \sqrt{2[(s+b)ln(1+s/b)-s]} \le 2$, where
$s$ and $b$ stand for the number of signal and background events, respectively.
It means that if no excess of events are recorded in HL-LHC, the colored
regions would be the remaining allowed regions. 
We clearly see significant gains using the 3CNN analysis for all four-types of 2HDMs. The 3CNN analysis has the potential to provide stronger constraints 
than the baseline method and BDT.

The 3CNN analysis shows stronger background discrimination power and thus provides a better coverage of parameter space at the HL-LHC.
Therefore, we focus on the 3CNN analysis in the following and combine with the current constraints from the Higgs-signal strengths obtained at the LHC and direct searches at high energy colliders. The current constraints are calculated from the public code \textsc{HiggsBounds}-v5.10.2 and \textsc{HiggsSignals}-v2.6.2. In the \textsc{HiggsBounds}-v5.10.2, 
it includes all processes at LEP, Tevatron, and LHC and determines
which is the most sensitive channel and whether the point is still allowed or not at the 95\% CL. In the \textsc{HiggsSignals}-v2.6.2, it gives the $\chi^2$ output for 111 Higgs observables~\cite{ATLAS:2018jvf,ATLAS:2018xbv,ATLAS:2018ynr,ATLAS:2020rej,CMS:2018hnq,CMS:2018nak,CMS-PAS-HIG-19-001,CMS-PAS-HIG-19-002}. Since there are six model parameters, the number of degrees of freedom is 105. We require the $p$-value to be larger than 0.05, corresponding to 2$\sigma$.

\begin{figure}[h!]
% \begin{center}
\centering
    \begin{subfigure}{0.49\textwidth}
    \centering
    \includegraphics[width=1.\columnwidth]{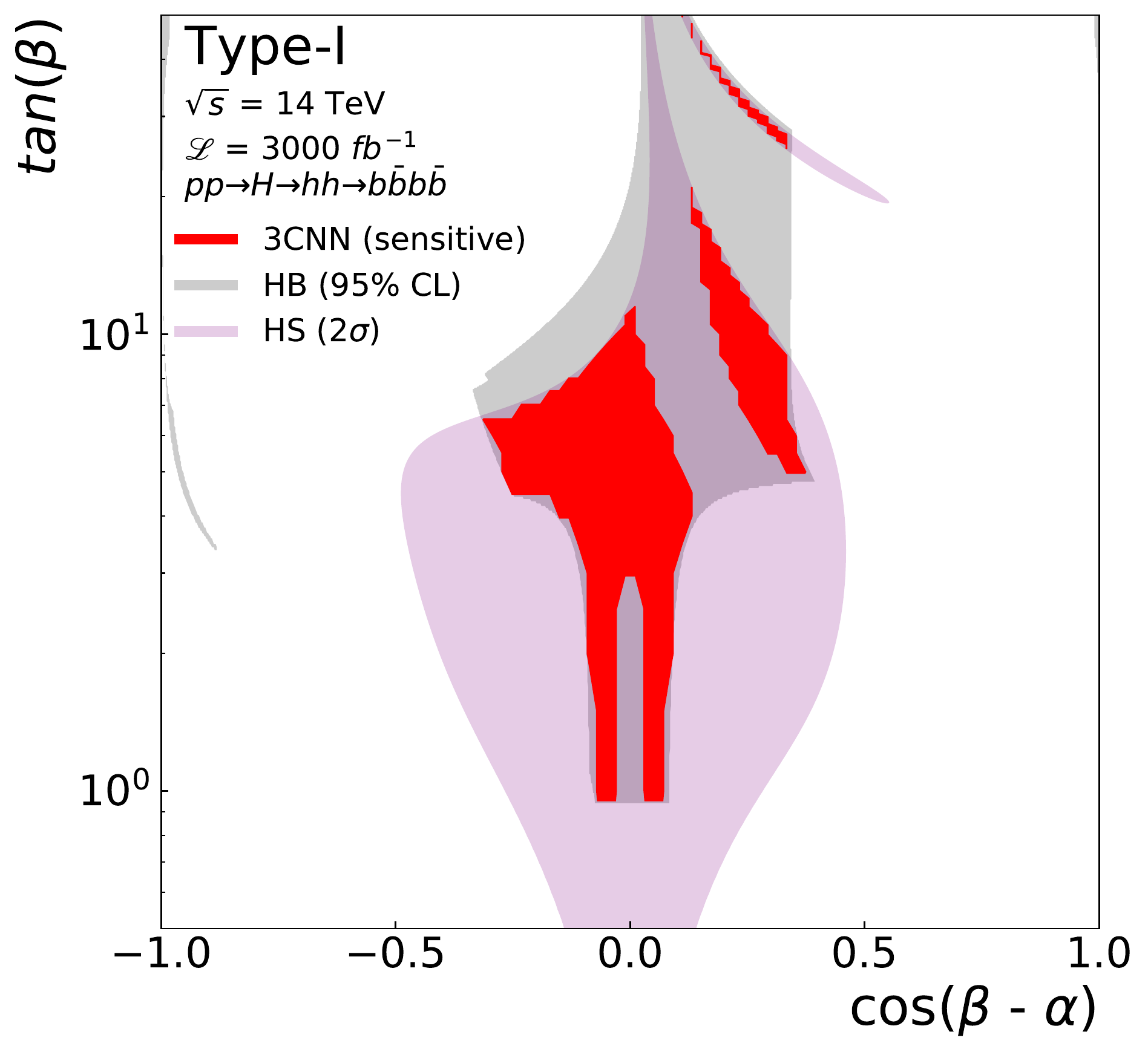}  
    \end{subfigure}
    \begin{subfigure}{0.49\textwidth}
    \centering
    \includegraphics[width=1.\columnwidth]{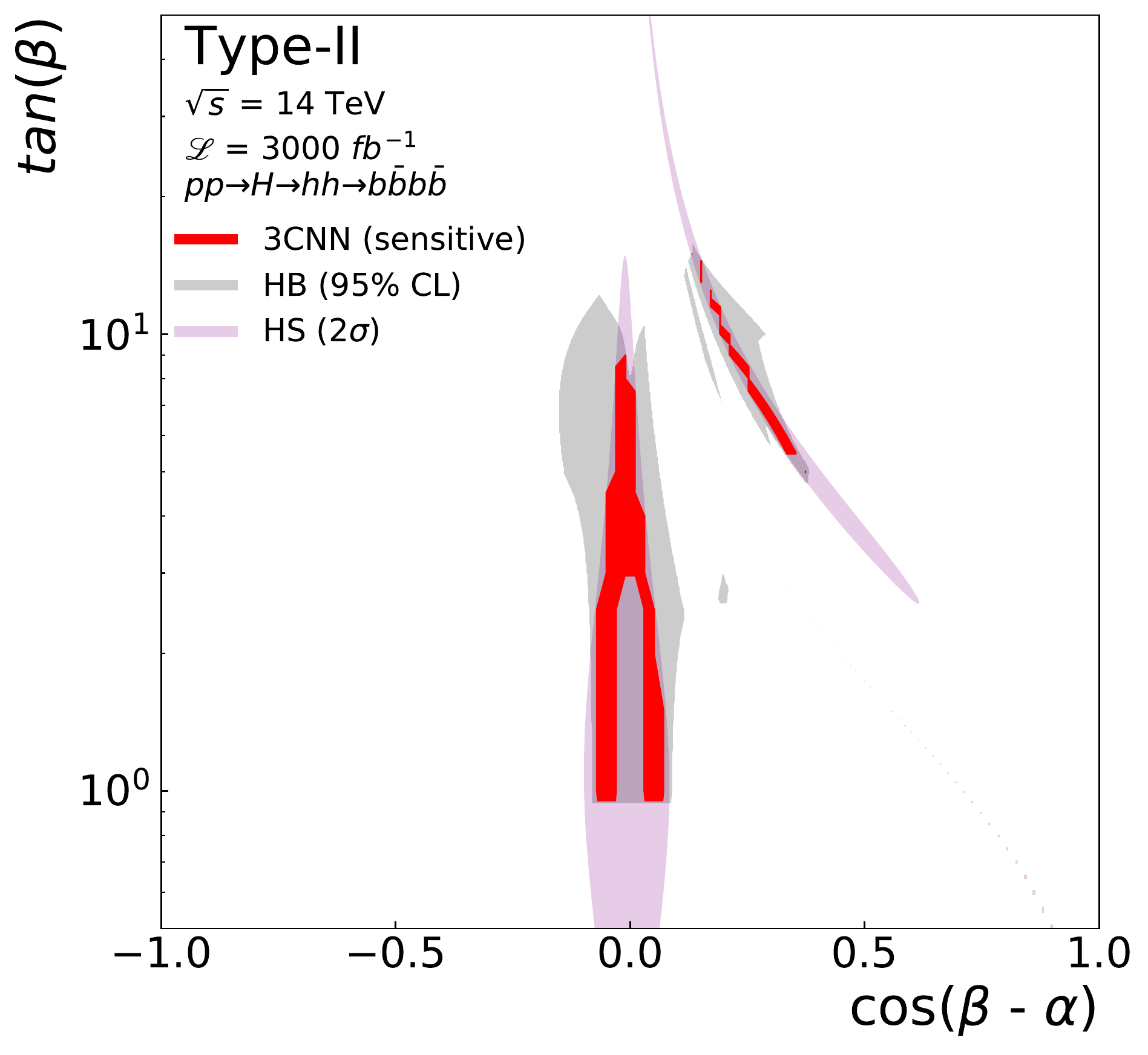}  
    \end{subfigure}
    \begin{subfigure}{0.49\textwidth}
    \centering
    \includegraphics[width=1.\columnwidth]{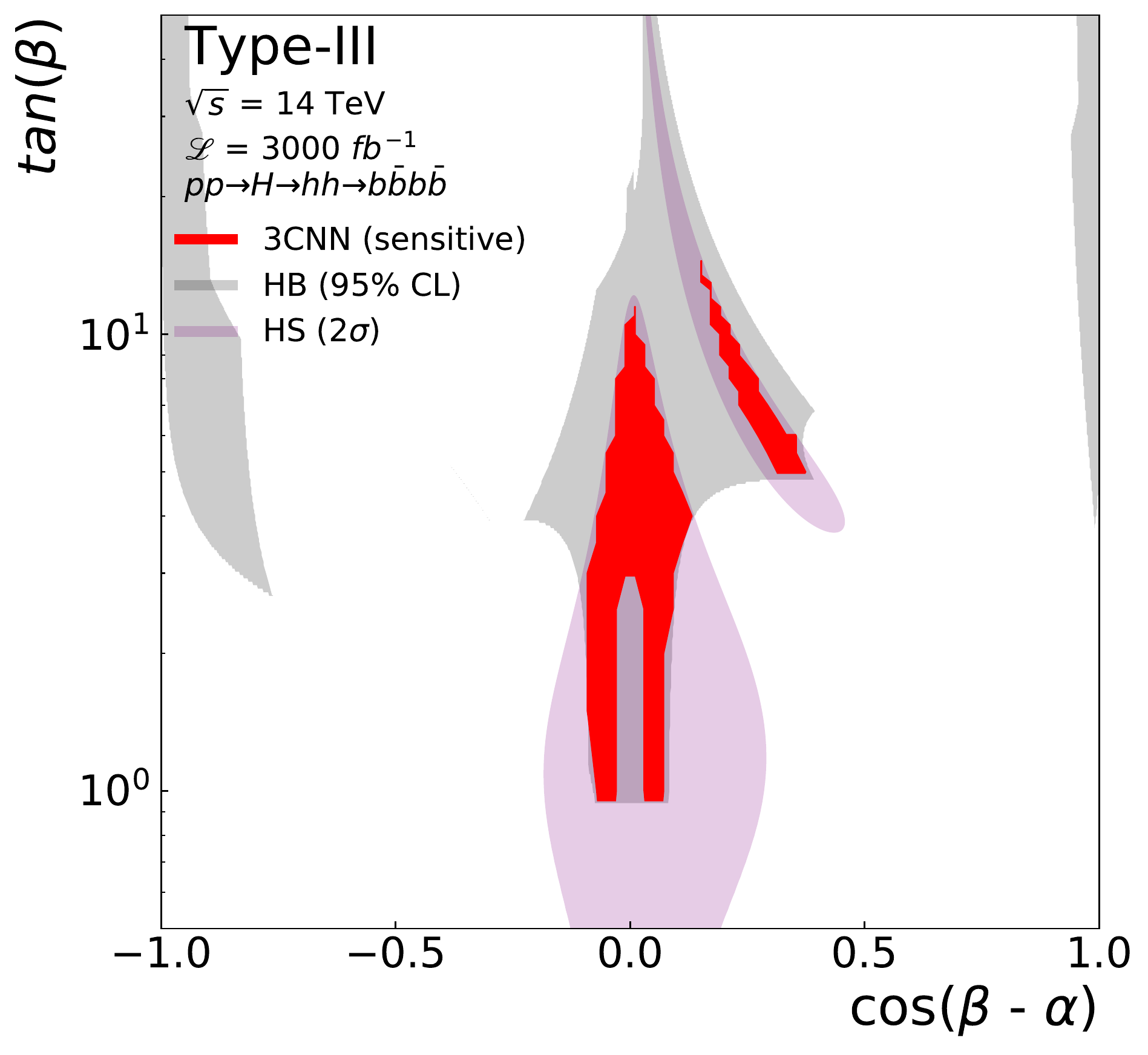}  
    \end{subfigure}
    \begin{subfigure}{0.49\textwidth}
    \centering
    \includegraphics[width=1.\columnwidth]{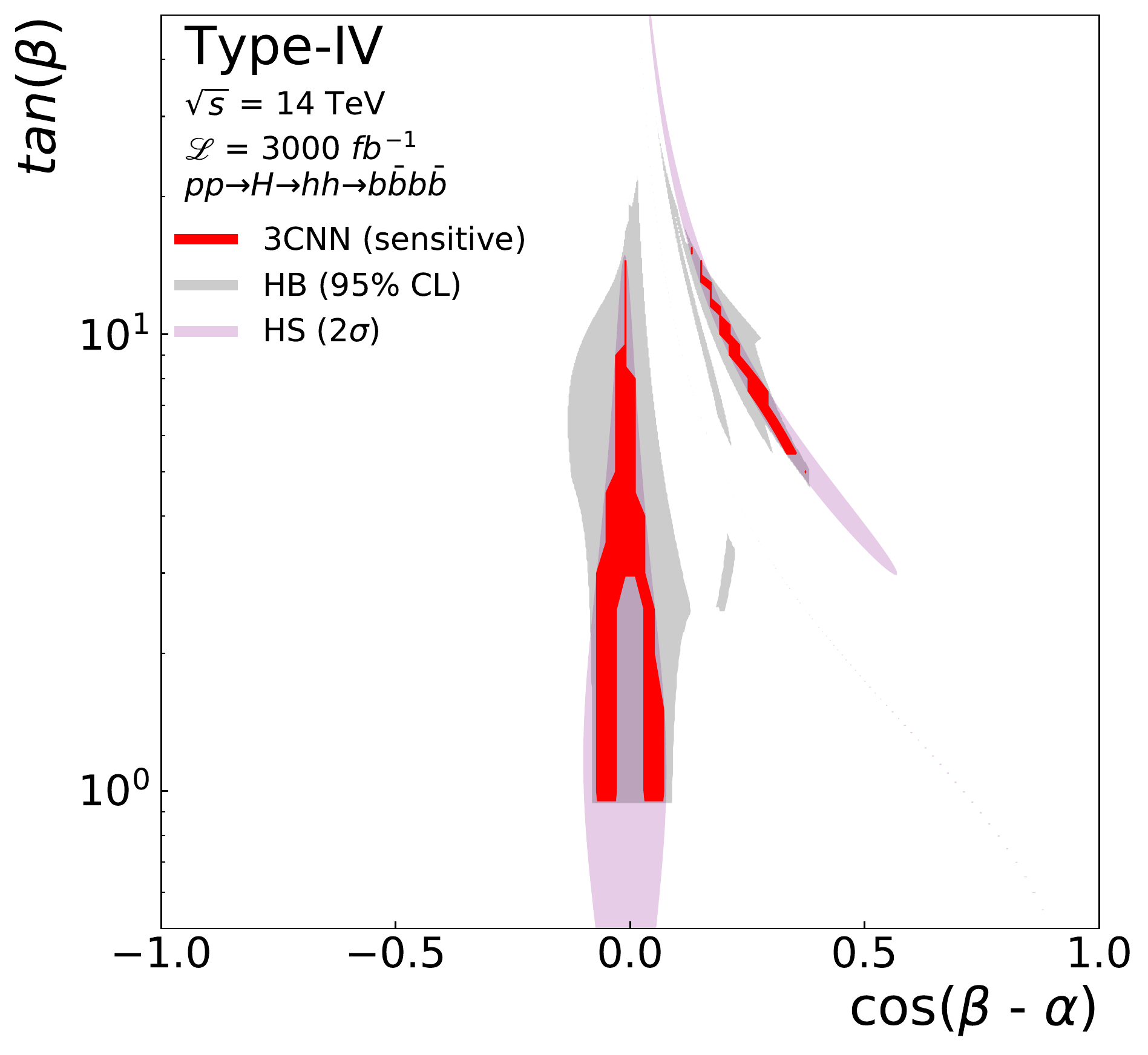}  
    \end{subfigure}
\caption{Sensitive regions in all four types of 2HDM for $M_A=M_{H^{\pm}}$ = 1000 GeV, $M_{12}^2$ = 400,000 $\text{GeV}^2$ in the ($\cos(\beta-\alpha)$, $\tan\beta$) plane. The gray area is the currently allowed area by direct searches at colliders from \texttt{HiggsBounds} at the 95\% CL. The purple area is due to the constraints from the SM-like Higgs-boson properties from \texttt{HiggsSignals} at 2 $\sigma$ level. The red region is the sensitive region (significance $>$ 2, where significance is $\sqrt{2[(s+b)ln(1+s/b)-s]}$,
and $s$ is the number of signal events and $b$ is the number of background events) passed the 3CNN analysis, where is still allowed in the current constraints at colliders.}
\label{fig:sensitivity_tb_cba}
% \end{center}
\end{figure}

\begin{figure}[h!]
% \begin{center}
\centering
    \begin{subfigure}{0.49\textwidth}
    \centering
    \includegraphics[width=1.\columnwidth]{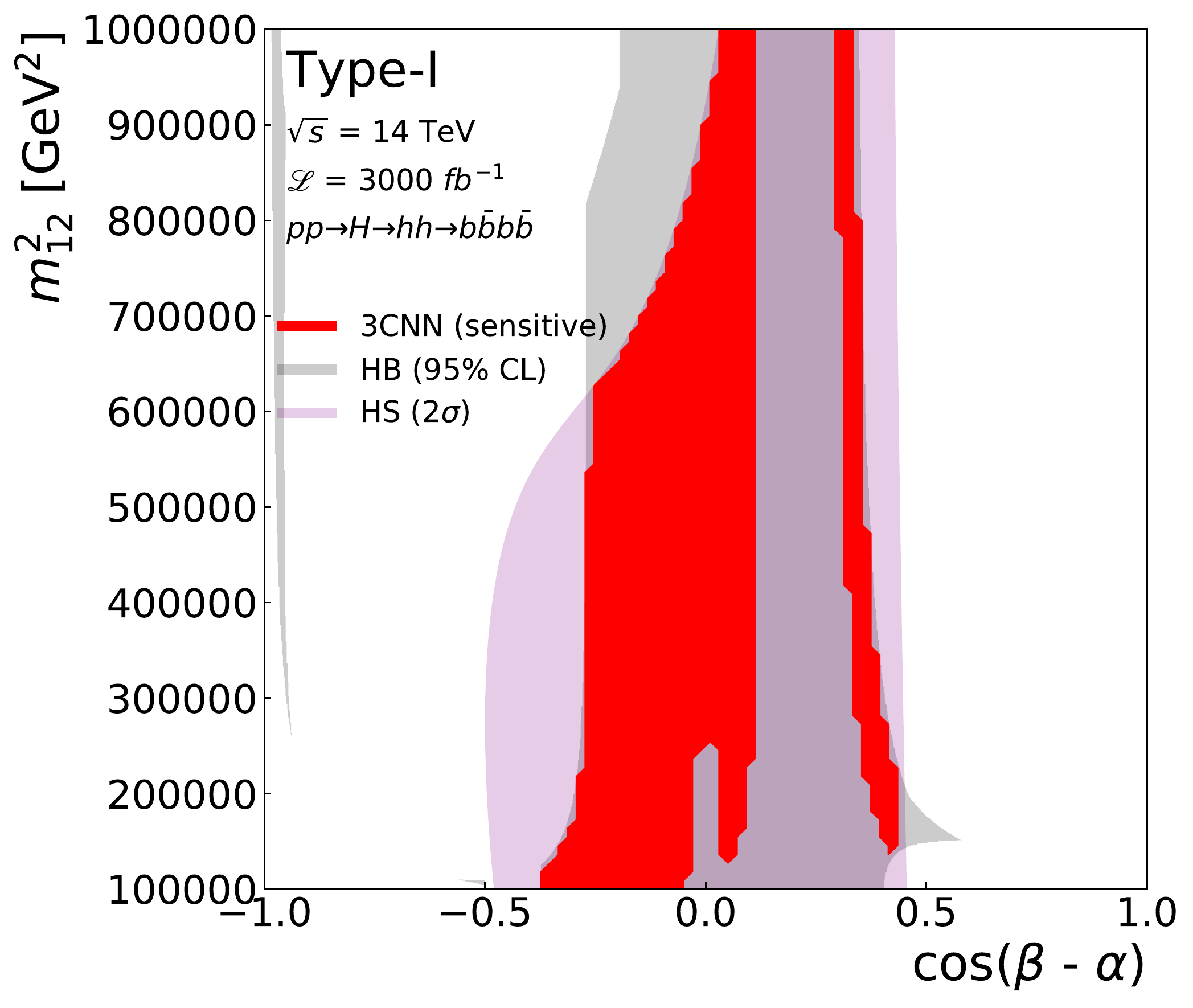}  
    \end{subfigure}
    \begin{subfigure}{0.49\textwidth}
    \centering
    \includegraphics[width=1.\columnwidth]{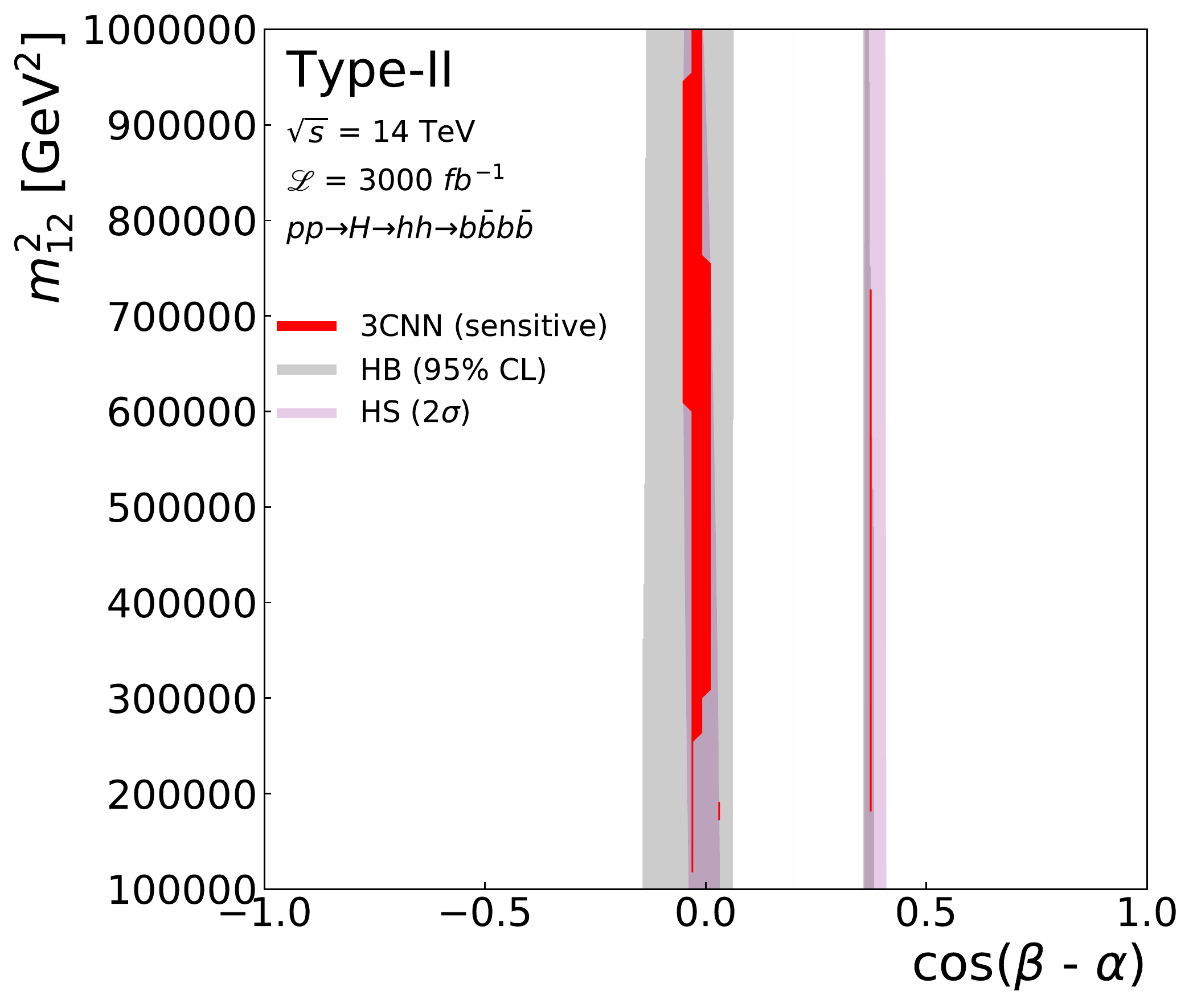}  
    \end{subfigure}
    \begin{subfigure}{0.49\textwidth}
    \centering
    \includegraphics[width=1.\columnwidth]{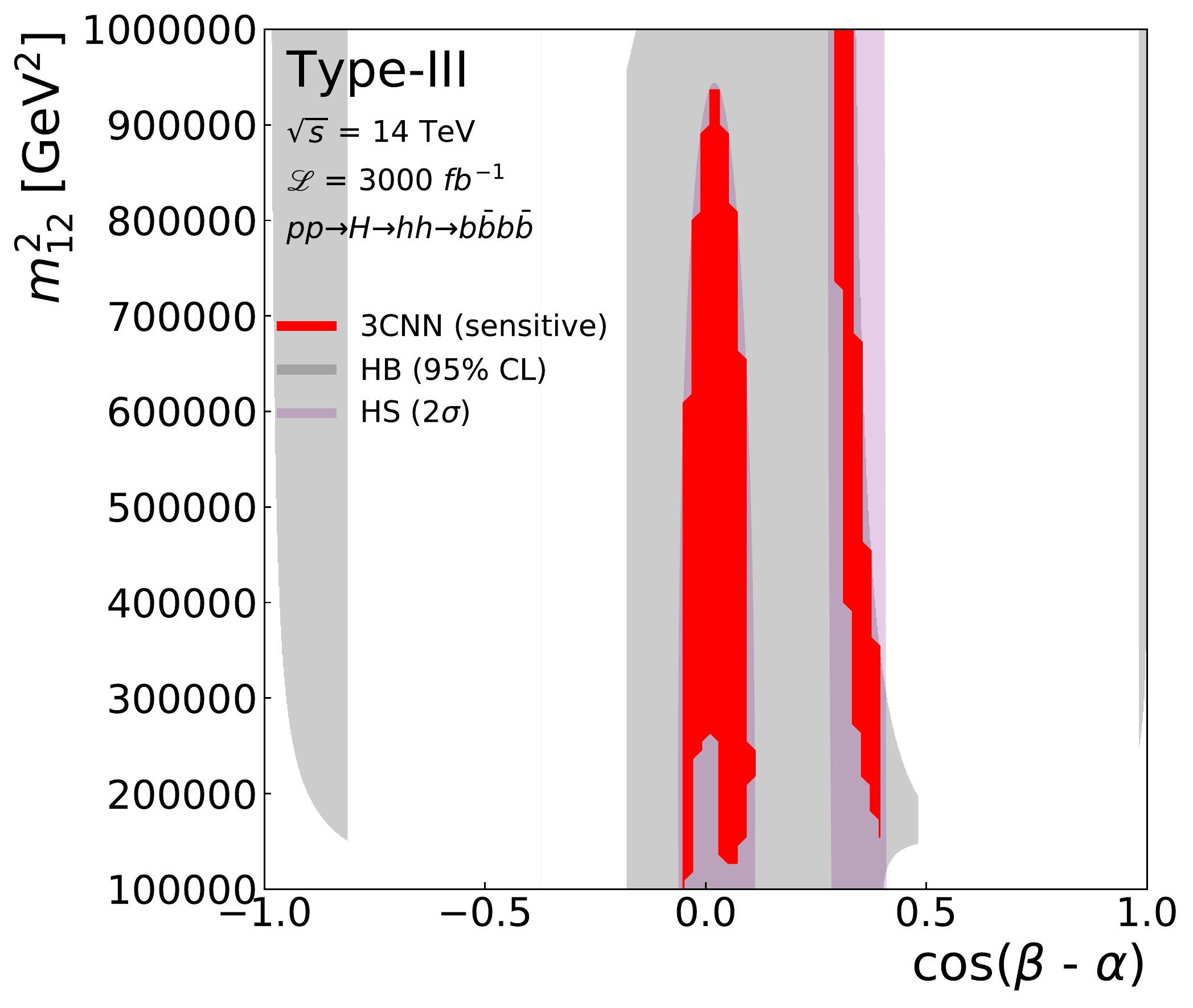}  
    \end{subfigure}
    \begin{subfigure}{0.49\textwidth}
    \centering
    \includegraphics[width=1.\columnwidth]{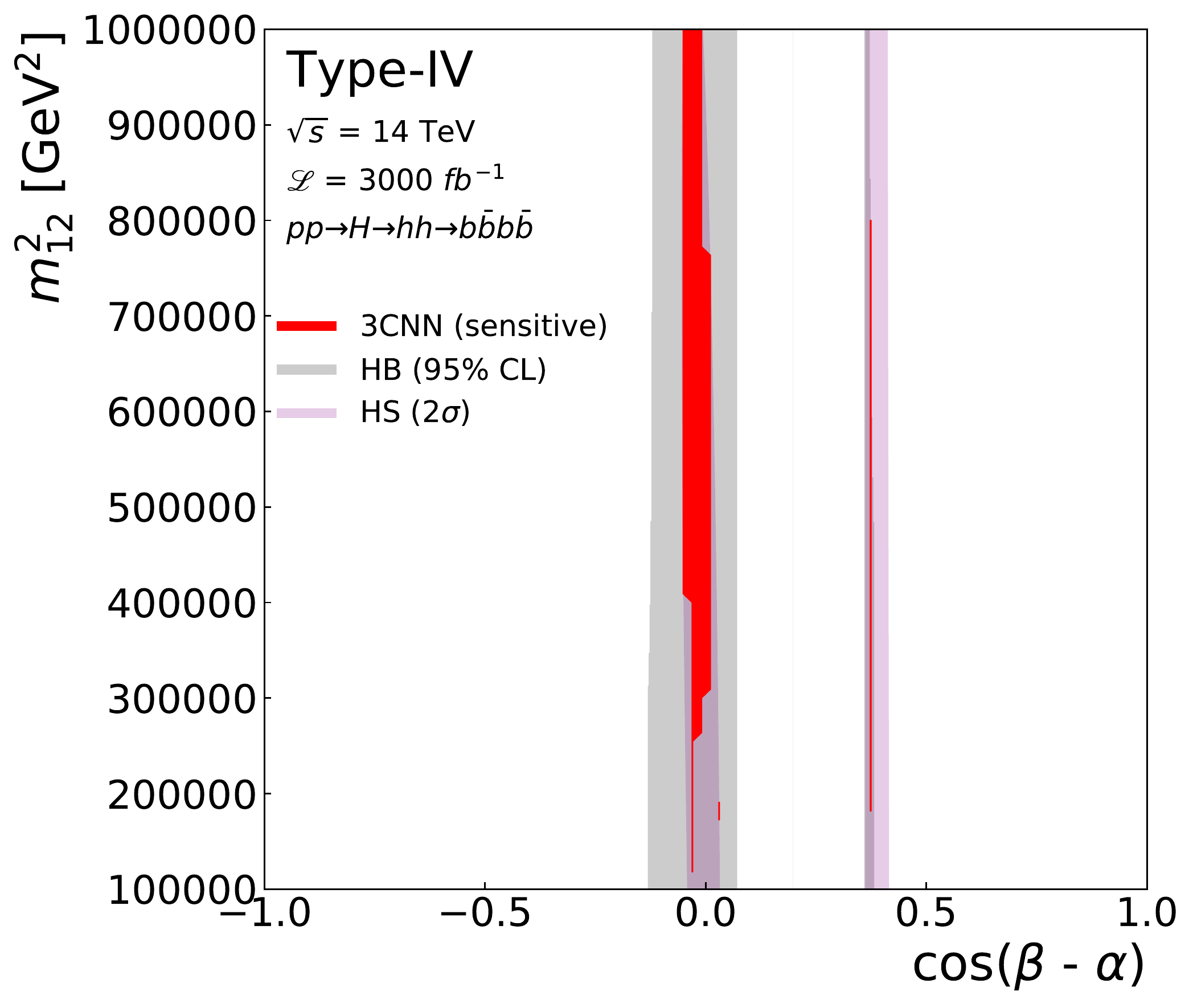}  
    \end{subfigure}
\caption{Sensitive region in all four types of 2HDM for $M_A=M_{H^{\pm}}$ = 1000 GeV, $\tan\beta$ = 5 in the ($\cos(\beta-\alpha)$, $M_{12}^2$) plane. The gray area is the current allowed area by direct searches at colliders from \texttt{HiggsBounds} at the 95\% CL. The purple area is due to the constraints from the SM-like Higgs-boson properties from \texttt{HiggsSignals} at 2 $\sigma$ level. The red region is the sensitive region (significance $>$ 2, significance is $\sqrt{2[(s+b)ln(1+s/b)-s]}$, where $s$ is the number of signal events 
and $b$ is the number of background events) passed the 3CNN analysis, where is still allowed in the current constraints at colliders.}
\label{fig:sensitivity_m12s_cba}
% \end{center}
\end{figure}

In the Fig.\ref{fig:sensitivity_tb_cba} and Fig.\ref{fig:sensitivity_m12s_cba},
we present the sensitivity region (red) with significance $z>2$ that is still allowed under current constraints and can be covered by the 3CNN at the 14 TeV HL-LHC in the ($\cos(\beta-\alpha)$, $\tan\beta$) plane and ($\cos(\beta-\alpha)$, $m_{12}^2$) plane, respectively. 
Note that the gray area is the currently allowed region by direct searches at colliders from \texttt{HiggsBounds} at the 95\% CL and the purple area is the 
allowed region from the SM-like Higgs-boson properties given by  \texttt{HiggsSignals} at 2 $\sigma$ level. We can regard the overlapping 
regions of the gray and purple areas as the currently allowed parameter space.
Note that the overlapping regions can be separated into 
(i) near the alignment limit and (ii) the wrong-sign Yukawa region.
In all 4 types of 2HDM, we clearly see that the 3CNN can cover a large area
of the overlapping regions.

In the ($\cos(\beta-\alpha)$, $\tan\beta$) plane of 
Fig.~\ref{fig:sensitivity_tb_cba}, we fix $M_A=M_{H^{\pm}}$ = 1000 GeV, and $M_{12}^2$ = 400,000 $\text{GeV}^2$.
In all 4 types of 2HDM, the red region is the sensitive region where the significance is larger than 2.
We can see that large areas (red) in the overlapping region of the gray and purple areas allowed by both \texttt{HiggsSignals} and \texttt{HiggsBounds} 
can be covered by 3CNN, indicating that the process 
$pp \to H \to h h \to 4b$ can test a large chunk of parameter space at the HL-LHC.
We notice that the sensitive regions lie close to the alignment limit, $\cos(\beta-\alpha)=0$, and in the wrong-sign region in all four types. 
Around the alignment limit, \texttt{HiggsBounds} restricts $\tan(\beta)$ 
to be larger than 1 and the 3CNN analysis shows that the region with 
$\tan(\beta) \le 10 $ is still sensitive. On the other hand, around the wrong sign Yukawa region, the most severe constraint comes from \texttt{HiggsSignals}. The 3CNN analysis also indicates that it can cover  a quite sizable area in the wrong sign Yukawa region.

On the other hand, in the ($\cos(\beta-\alpha)$, $m_{12}^2$) plane 
of Fig.~\ref{fig:sensitivity_m12s_cba}, we fix 
$M_A=M_{H^{\pm}}$ = 1000 GeV and $\tan\beta$ = 5.
The 3CNN analysis indicates the sensitivity in the $m_{12}^2$ interval around alignment limit and along the wrong sign Yukawa region.

\section{Conclusions}\label{sec:conclusion}
In this study, we have employed a modern deep-learning approach to improve the search for Higgs boson pair production arising from resonant heavy Higgs enhancement 
in the the $b\bar{b}b\bar{b}$ final state in the framework of two-Higgs-doublet models at the HL-LHC. The resonance production channel plays an important role in probing the structure of the EWSB sector. Using our approach, we have pointed out that the gluon-fusion process $pp \to H \to h h \to 4b$ at the HL-LHC can further probe the currently allowed parameter space in the Types I to IV of 2HDM's.

The 3CNN architecture in this work is built upon the proposal from Ref.~\cite{Lin:2018cin,Chung:2020ysf}. This architecture has 2-class outputs for the signal and background, and contains one stream acting on global event information, and the other two streams acting on information from the leading and subleading jets. This approach is amenable to visualizations that can provide some insights into what the neural network is using for event classification.

We interpret the signal-background discrimination based on our simulations at 14 TeV HL-LHC in the two-Higgs-doublet models' framework. Figures  \ref{fig:sensitivity_tb_cba} and \ref{fig:sensitivity_m12s_cba} illustrate our scanning in the parameter space of the 2HDM’s for the sensitivity coverage at the HL-LHC, as well as the current restriction on the parameter space due to \textsc{HiggsSignals} and \textsc{HiggsBounds}. We find that there is sizeable sensitive parameter space covered by the 3CNN analysis.

In summary, we employ the 3CNN architecture to incorporate both local and global information for the signal and background identification. Additionally, we have studied the conventional cut-based approach and a boosted decision tree. The conventional cut-based approach does not give enough significance to the signal even at HL-LHC. The BDT is effective but is less potent than the neural network. We have shown that the 3CNN can significantly enhance the significance of the signal at HL-LHC and allows us to probe sensitive parameter space in the currently allowed region. This work is flexible to implement in other Higgs-pair production channels with hadronic or semi-hadronic final state and may be able to enrich the sensitivity of the signal at the HL-LHC.

\begin{acknowledgments}
We thank SooJin Lee for help with \textsc{HiggsSignals} and \textsc{HiggsBounds}, also thank Professor Benjamin Nachman and Professor Chih-Ting Lu for their valuable comments on the manuscript. K.C. and Y. C. were supported by MoST with grant nos. MoST-110-2112-M-007-017-MY3. S.C.H was supported by the National Science Foundation under Grant No. 2110963.
\end{acknowledgments}

\clearpage
% %%%%%%%%%%%%%%%%%%%%%%%%%%%%%%%%%%%%%%%%%%%%%%%%%%%%%%%%%%%%%%%%%%%%%%%%%%%%%%%%s
\bibliographystyle{jhep}
\bibliography{references,HEPML}
% \bibliography{references,HEPML,another}

\end{document}